\documentclass[final, 10pt]{IEEEtran}
\IEEEoverridecommandlockouts

\ifCLASSOPTIONpeerreview
\renewcommand{\baselinestretch}{2}
\fi

\usepackage[cmex10]{amsmath} 
\usepackage{amssymb}
\usepackage{bm}
\usepackage{mathtools}
\usepackage{cite}
\usepackage{xcolor}
\usepackage{enumerate}
\usepackage{booktabs} 
\usepackage{multirow}

\usepackage{graphicx}
\graphicspath{{figs/}{pdf/}{jpg/}{../pdf/}}

\newcommand{\figref}[1]{Fig.\,\ref{#1}}
\newcommand{\secref}[1]{Section~\ref{#1}}

\DeclarePairedDelimiterX\abs[1]{\lvert}{\rvert}{#1}
\DeclarePairedDelimiterX\brak[1]{\lbrace}{\rbrace}{#1}
\DeclarePairedDelimiterX\cardinality[1]{\lvert}{\rvert}{#1}
\DeclarePairedDelimiterX\innerp[2]{\langle}{\rangle}{#1,#2}
\DeclarePairedDelimiterX\norm[1]{\lVert}{\rVert}{#1}
\DeclarePairedDelimiterX\paren[1]{(}{)}{#1} 
\DeclarePairedDelimiterX\parn[1]{(}{)}{#1}
\DeclarePairedDelimiterX\sequ[1]{\lbrace}{\rbrace}{#1}
\DeclarePairedDelimiterX\set[1]{\lbrace}{\rbrace}{#1}
\DeclarePairedDelimiterX\sqrb[1]{[}{]}{#1}
\DeclarePairedDelimiterX\coeff[1]{(}{)}{#1}

\newcommand{\lsph}{L^2(\untsph)}

\newcommand{\dfn}{\triangleq}
\newcommand{\boldhat}[1]{\widehat{\bm{#1}}}

\newcommand{\unit}[1]{\boldhat{#1}}
\newcommand{\conj}[1]{\overline{#1}} 

\newcommand{\reals}{\mathbb{R}} 

\newcommand{\untsph}{\mathbb{S}^{2}} 

\newcommand{\intsph}{\int_{\untsph}}

\usepackage{amsthm,url}
\usepackage{subfig} 

\newcommand{\A}{\textbf{A}}
\newcommand{\shc}[3]{({#1})_{#2}^{#3}}
\newcommand{\bv}[1]{\boldsymbol{#1}}

\newcommand{\summr}{\smashoperator[r]{\sum_{m=-\ell}^{\ell}}\,}
\newcommand{\summ}{\smashoperator{\sum_{m=-\ell}^{\ell}}\,}
\newcommand{\summp}{\smashoperator{\sum_{m'=-\ell}^{\ell}}\,}
\newcommand{\matlab}{\texttt{MATLAB}}

\newtheorem{definition}{Definition}

\renewcommand{\IEEEQED}{\IEEEQEDopen}

\newcommand{\AuthorOne}{Yibeltal~F.~Alem, {\em{Student Member, IEEE}}}
\newcommand{\AuthorTwo}{Zubair~Khalid, {\em{Member, IEEE}}}
\newcommand{\AuthorThree}{Rodney A. Kennedy, {\em{Fellow, IEEE}}}

\newcommand{\ThankOne}{The authors are with the Research School of Engineering, College of Engineering and Computer Science, The
Australian National University, Canberra, ACT 2601, Australia.
Emails: \{yilbeltal.alem, zubair.khalid, rodney.kennedy\}@anu.edu.au}

\newcommand{\ThankTwo}{This work was supported by the Australian Research Council's Discovery Projects funding scheme (Project no. DP150101011).}

\begin{document}

\title{Spherical Harmonic Expansion of Fisher-Bingham Distribution and 3D Spatial Fading Correlation for Multiple-Antenna Systems}

\author{\authorblockN{\AuthorOne, \AuthorTwo \: and \AuthorThree
\thanks{\ThankOne. \ThankTwo}}}

\maketitle

\begin{abstract}
This paper considers the 3D spatial fading correlation (SFC) resulting
from an angle-of-arrival (AoA) distribution that can be modelled by a mixture
of Fisher-Bingham distributions on the sphere.  By deriving a closed-form
expression for the spherical harmonic transform for the component Fisher-Bingham
distributions, with arbitrary parameter values, we obtain a closed-form
expression of the 3D-SFC for the mixture case. The 3D-SFC expression is
general and can be used in arbitrary multi-antenna array geometries and is
demonstrated for the cases of a 2D uniform circular array in the horizontal
plane and a 3D regular dodecahedral array.  In computational aspects, we
use recursions to compute the spherical harmonic coefficients and give
pragmatic guidelines on the truncation size in the series representations to
yield machine precision accuracy results. The results are further corroborated
through numerical experiments to demonstrate that the closed-form expressions
yield the same results as significantly more computationally expensive numerical
integration methods.
\end{abstract}

\begin{IEEEkeywords}
Fisher-Bingham distribution, spherical harmonic expansion, spatial correlation,
MIMO, angle of arrival~(AoA).
\end{IEEEkeywords}


\section{Introduction}

The Fisher-Bingham distribution, also known as the Kent distribution,
belongs to the family of spherical distributions in directional statistics
\cite{Kent:1982}. It has been used in applications in a wide range of disciplines
for modelling and analysing directional data. These applications
include sound source
localization~\cite{Leong:1998,Langendijk:2001}, joint set identification~\cite{Peel:2001},
modelling protein structures~\cite{Kent:2005}, 3D beamforming~\cite{Christou:2008},
classification of remote sensing data \cite{Lunga:2011}, modelling the
distribution of AoA in wireless
communication~\cite{Mammasis:2008,Mammasis:2009}, to name a few.

In this work, we focus our attention to the use of Fisher-Bingham distribution
for modelling the distribution of angle of arrival~(AoA)~(also referred as the
distribution of scatterers) in wireless communication and computation of
spatial
fading correlation~(SFC) experienced between elements of multiple-antenna array
systems. Modelling the distribution of scatterers and characterising the spatial
correlation of fading channels is a key factor in evaluating the performance of
wireless communication systems with multiple antenna
elements~\cite{Salz:1994,Fang:2000,Shiu:2000,Abdi:2002,Tsai:2004,Kennedy:2007}.
It has been an active area of research for the past two decades or so and a
number of spatial correlation models and closed-form expressions for evaluating
the SFC function have been developed in the existing
literature~(e.g.,~\cite{Salz:1994,Kalkan:1997,Pedersen:1997,Vaughan:1998,
Abdi:2000,Tsai:2002,Teal:2002,Yong:2005,Mammasis:2010,Lee:2012,Kennedy:2013b}).

The elliptic~(directional) nature of the Fisher-Bingham distribution offers
flexibility in modelling the distribution of normalized power or AoA of the
multipath components for most practical scenarios. Exploiting this fact, a 3D
spatial correlation model has been developed in~\cite{Mammasis:2009}, where the
distribution of AoA of the multipath components is modelled by a positive linear
sum of Fisher-Bingham distributions, each with different parameters. Such
modelling has been shown to be useful in a sense that it fits well with the
multi-input multi-output (MIMO) field data and allows the evaluation of the
SFC as a function of the angular spread, ovalness parameter, azimuth, elevation,
and prior contribution of each cluster. Although the SFC function presented in
\cite{Mammasis:2009} is general in a sense that it is valid for any arbitrary
antenna array geometry, it has not been expressed in closed-form and has been
only evaluated using numerical integration techniques, which can be
computationally intensive to produce sufficiently accurate results. If the
spherical harmonic expansion of the Fisher-Bingham distribution is given in a
closed-form, the SFC function can be computed analytically using the spherical
harmonic expansion of the distribution of AoA of the multipath
components~\cite{Teal:2002}. To the best of our knowledge, the spherical
harmonic expansion of Fisher-Bingham distribution has not been derived in the
existing literature.

In the current work, we present the spherical harmonic expansion of
Fisher-Bingham distribution and a closed-form expression that enables the
analytic computation of the spherical harmonic coefficients. We also address
the computational considerations required to be taken into account in the
evaluation of the proposed closed-form. Using the proposed spherical harmonic
expansion of the Fisher-Bingham distribution, we also formulate the SFC function
experienced between two arbitrary points in 3D-space for the case when the
distribution of AoA of the multipath components is modelled by a mixture
(positive linear
sum) of Fisher-Bingham distributions. The SFC presented here is general
in a sense that it is expressed as a function of arbitrary points in 3D-space
and therefore can be used to compute spatial correlation for any 2D and 3D
antenna array geometries. Through numerical analysis, we also validate the
correctness of the proposed spherical harmonic expansion of Fisher-Bingham
distribution and the SFC function. In this paper, our main objective is to
employ the proposed spherical harmonic expansion of the Fisher-Bingham
distribution for computing the spatial correlation. However, we expect that the
proposed spherical harmonic expansion can be useful in various applications
where the Fisher-Bingham distribution is used to model and analyse directional
data~(e.g.,~\cite{Leong:1998,Langendijk:2001,Peel:2001,Kent:2005,Christou:2008,
Lunga:2011}).

The remainder of the paper is structured as follows. We review the mathematical
background related to signals defined on the 2-sphere and spherical harmonics in
\secref{sec:preliminaries}. In \secref{sec:problem}, we define the
Fisher-Bingham distribution and present its application in modelling 3D spatial
correlation. We present the spherical harmonic expansion of the Fisher-Bingham
distribution and analytical formula for computing the spherical harmonic
coefficients in \secref{sec:expansion}, where we also address computational
issues. We derive a closed-form expression for the 3D SFC between two arbitrary
points in 3D-space when the AoA of an incident signal follows the Fisher-Bingham
probability density function~(pdf) in \secref{sec:application}. In
\secref{sec:analysis}, we carry out a numerical validation of the proposed
results and provide examples of the SFC for uniform circular array (2D) and
regular dodecahedron array (3D) antenna elements. Finally, the concluding
remarks are made in \secref{sec:conclusion}.


\section{Mathematical Preliminaries}\label{sec:preliminaries}

\subsection{Signals on the 2-Sphere}

We consider complex valued square integrable functions defined on the 2-sphere,
$\mathbb{S}^2$. The set of such functions forms a Hilbert space, denoted by
$\lsph$, that is is equipped with the inner product defined for two functions
$f$ and $g$ defined on $\mathbb{S}^2$ as~\cite{Kennedy-book:2013}
\begin{equation}
\label{eq:L2-ip}
    \innerp{f}{g}\dfn \intsph f(\unit{x}) \conj{g(\unit{x})}\,ds(\unit{x}),
\end{equation}
which induces a norm $\norm{f}\dfn\innerp{f}{f}^{1/2}$. Here,
$\overline{(\cdot)}$ denotes the complex conjugate operation and
$\unit{x}\dfn[\sin\theta\cos\theta,\sin\theta\sin\phi,
\cos\theta]^T\in\untsph\subset \mathbb{R}^{3}$ represents a point on the
2-sphere, where $[\cdot]^T$ represents the vector transpose, $\theta \in
[0, \pi]$ and $\phi\in [0, 2\pi)$ denote the co-latitude and
longitude respectively and $ds(\unit{x})=\sin\theta\,d\theta\,d\phi$ is the
surface measure on the 2-sphere. The functions with finite energy~(induced norm)
are referred as signals on the sphere.

\subsection{Spherical Harmonics}
Spherical harmonics serve as orthonormal basis functions for the representation
of functions on the sphere and are defined for integer degree $\ell\ge 0$
and integer order $|m|\le \ell$ as
\begin{equation*}
Y_{\ell}^{m}(\unit{x}) \equiv	Y_{\ell}^{m}(\theta,\phi) \dfn
		N_\ell^m
		P_{\ell}^{m}(\cos\theta)e^{im\phi},
\end{equation*}
with
\begin{equation}
N_{\ell}^{m} \dfn \sqrt{\frac{2{\ell}+1}{4\pi}\frac{({\ell}-m)!}{({\ell}+m)!}},
\end{equation}
is the normalization factor such that $\innerp[\big]{Y_{\ell}^m}{ Y_{p}^{q}}=\delta_{\ell, p} \delta_{m,q}$, where $\delta_{m,q}$ is the Kronecker delta function: $\delta_{m,q} = 1$ for $m=q$ and is zero otherwise.
$P_{\ell}^{m}(\cdot)$ denotes the associated Legendre polynomial of degree $\ell$ and order $m$~\cite{Kennedy-book:2013}. We also note the following relation for associated Legendre polynomial
\begin{equation}\label{Eq:Plm_Wignerd}
	P_{\ell}^m(\cos\theta)
		= \sqrt{\frac{(\ell+m)!}{(\ell-m)!}}\,d^{\ell}_{m,0}(\theta),
\end{equation}
where $d^{\ell}_{m,m'}(\cdot)$ denotes the Wigner-$d$ function of degree $\ell$ and orders $m$ and $m'$~\cite{Kennedy-book:2013}.

By completeness of spherical harmonics, any finite energy function
$f(\unit{x})$ on the 2-sphere can be expanded as
\begin{equation}\label{eq:f_expansion}
	f(\unit{x}) = \sum_{\ell=0}^{\infty} \summr
		 \shc{f}{\ell}{m}  \,Y_{\ell}^{m}(\unit{x}),
\end{equation}
where $\shc{f}{\ell}{m}$ is the spherical harmonic coefficient given by
\begin{equation}
\shc{f}{\ell}{m}=\intsph f(\unit{x})\conj{Y_{\ell}^{m}(\unit{x})}\,ds(\unit{x}).
\end{equation}
The signal $f$ is said to be band-limited in the spectral domain at degree $L$
if $\shc{f}{\ell}{m} = 0$ for $\ell>L$. For a real-valued function $f(\unit{x})$,
we also note the relation
\begin{equation}\label{Eq:conj_symmetry}
 \shc{f}{\ell}{-m}=(-1)^m\overline{\shc{f}{\ell}{m}},
\end{equation}
which stems from the conjugate symmetry property of spherical harmonics
\cite{Kennedy-book:2013}.

\subsection{Rotation on the Sphere}
The rotation group SO(3) is characterized by Euler angles
$(\varphi,\vartheta,\omega)\in \text{SO(3)}$, where $\varphi\in[0,2\pi)$,
$\vartheta\in[0,\pi]$ and $\omega\in[0,2\pi)$. We define a rotation operator
on the sphere $\mathcal{D}(\varphi,\vartheta,\omega)$ that
rotates a function on the sphere, according to $'zyz'$
convention, in the sequence of $\omega$ rotation around the $z$-axis,
$\vartheta$ rotation around the $y$-axis and $\varphi$ rotation around
$z$-axis. A rotation of a function $f(\theta,\phi)$ on sphere is given by
\begin{equation}\label{eq:rot_op}
 (\mathcal{D}(\varphi,\vartheta,\omega)f)(\unit
{x})\dfn f(\bv{R}^{-1}\unit{x}),
\end{equation}
where $\bv{R}$ is a $3\times3$ real orthogonal unitary matrix, referred as
rotation matrix, that corresponds to the rotation operator
$\mathcal{D}(\varphi,\vartheta,\omega)$ and is given by

\begin{equation}\label{eq:rot_matrix}
 \boldsymbol R = \boldsymbol R_z(\varphi)\boldsymbol R_y(\vartheta)\boldsymbol
R_z(\omega),
\end{equation}
where
\ifCLASSOPTIONonecolumn
\begin{equation*}
 \bv{R}_z(\varphi)=
 \begin{bmatrix}
  \cos \varphi & -\sin\varphi & 0\\
  \sin \varphi & \cos \varphi & 0\\
  0           & 0           & 1
 \end{bmatrix},\quad
 \bv{R}_y(\vartheta)=
 \begin{bmatrix}
  \cos \vartheta & 0 & \sin \vartheta\\
  0 & 1 & 0\\
  -\sin \vartheta    & 0     & \cos \vartheta
 \end{bmatrix}.
\end{equation*}
\else
\begin{equation*}
 \bv{R}_z(\varphi)=
 \begin{bmatrix}
  \cos \varphi & -\sin\varphi & 0\\
  \sin \varphi & \cos \varphi & 0\\
  0           & 0           & 1
 \end{bmatrix},
\end{equation*}

\begin{equation*}
 \bv{R}_y(\vartheta)=
 \begin{bmatrix}
  \cos \vartheta & 0 & \sin \vartheta\\
  0 & 1 & 0\\
  -\sin \vartheta    & 0     & \cos \vartheta
 \end{bmatrix}.
\end{equation*}
\fi
\noindent Here $\bv{R}_z(\varphi)$ and $\bv{R}_y(\vartheta)$ characterize individual rotations by $\varphi$ along $z$-axis and $\vartheta$ along
$y$-axis, respectively.


\section{Problem Formulation}\label{sec:problem}

\subsection{Fisher-Bingham Distribution on Sphere}

\begin{definition}[Fisher-Bingham Five-Parameter~(FB5) Distribution]
The Fisher-Bingham five-parameter~(FB5) distribution, also known as the Kent distribution, is a distribution on the sphere with probability density function~(pdf) defined as~\cite{Kent:1982,Kent:2005}
\begin{equation}\label{eq:kent_rotated}
g(\unit{x};\kappa,\unit{\mu},\beta,\A)=\frac{1}{C(\kappa,\beta\A)}e^{\kappa\unit{
\mu}^T\unit{x} + \unit{x}^T\beta\A\unit{x}},
\end{equation}
where $\A$ is a symmetric matrix of size
$3\times 3$ given by
\begin{equation}
 \A=(\unit{\eta}_1\unit{\eta}_1^T -
\unit{\eta}_2\unit{\eta}_2^T).
\end{equation}
\noindent Here $\unit{\mu}$, $\unit{\eta}_1$ and $\unit{\eta}_2$ are the unit
vectors~(orthonormal set) that denote the mean direction~(centre), major axis and minor axis of
the distribution, respectively, $\kappa\ge 0$ is the concentration parameter
that quantifies the spatial concentration of FB5 distribution around its mean
and $0\le\beta\le\kappa/2$ is referred to as the ovalness parameter that is a
measure of ellipticity of the distribution. In
\eqref{eq:kent_rotated}, the term $C(\kappa,\beta)$ denotes the normalization
constant, which ensures $\|g\|_2=1$ and is given by
\begin{equation}\label{eq:norm_constant}
 C(\kappa,\beta)=2\pi\sum_{r=0}^{\infty}\frac{\Gamma(r+1/2)}{\Gamma(r+1)}\beta^{
2r}(\kappa/2)^{-2r-1/2}I_{2r+1/2}(\kappa),
\end{equation}
where $I_r(\cdot)$ denotes the modified Bessel function of the first kind of
order $r$.
\end{definition}
\begin{figure*}[t]
    \centering
    \hspace{-6mm}
    \subfloat[$\kappa=25$, $\beta=10$]{
        \includegraphics[width=0.38\textwidth]{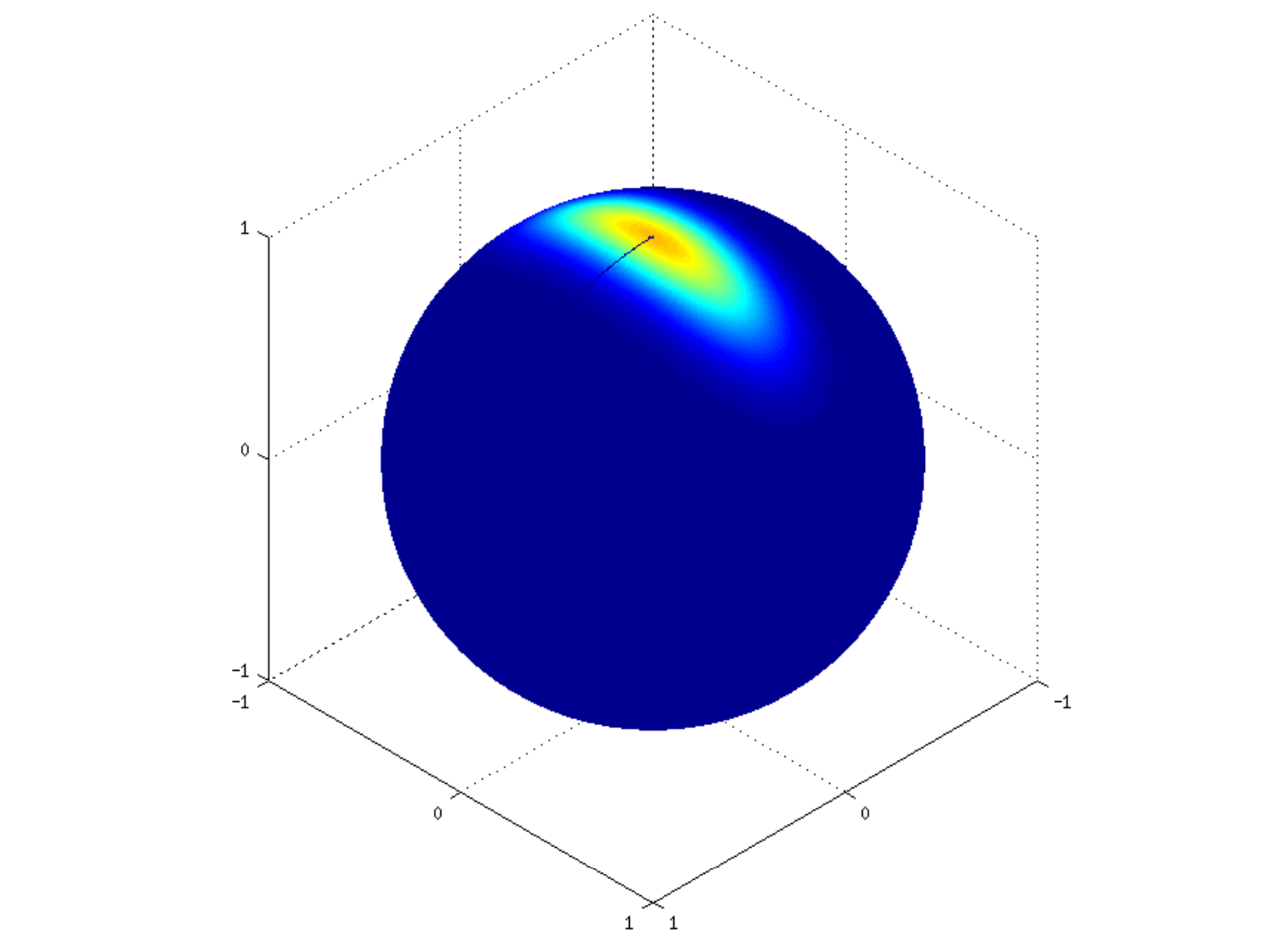}}\hfil
   \hspace{-20mm}
    \subfloat[$\kappa=100$, $\beta=10$]{
        \includegraphics[width=0.38\textwidth]{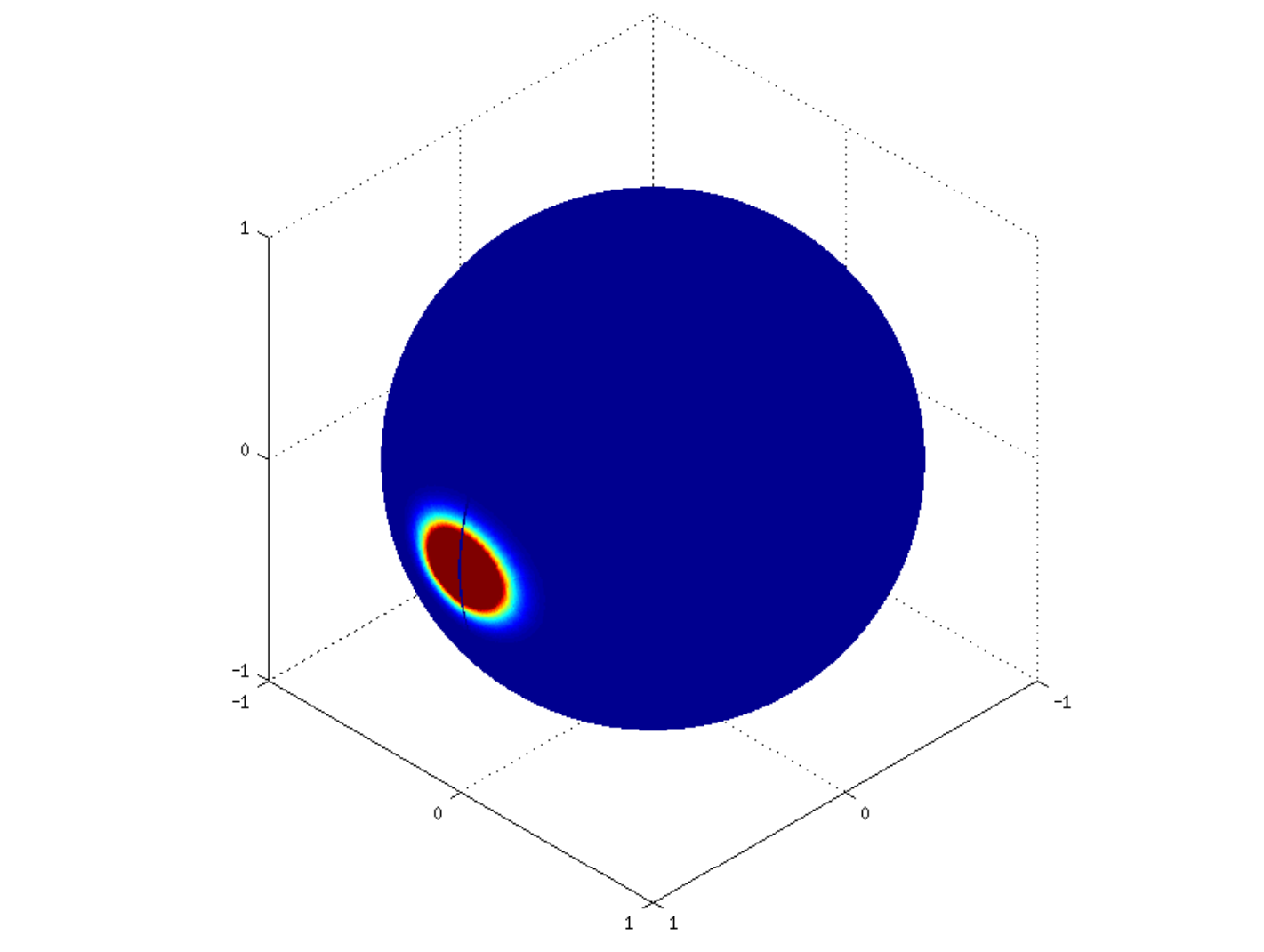}}\hfil
   \hspace{-20mm}
    \subfloat[$\kappa=100$, $\beta=49$]{
        \includegraphics[width=0.38\textwidth]{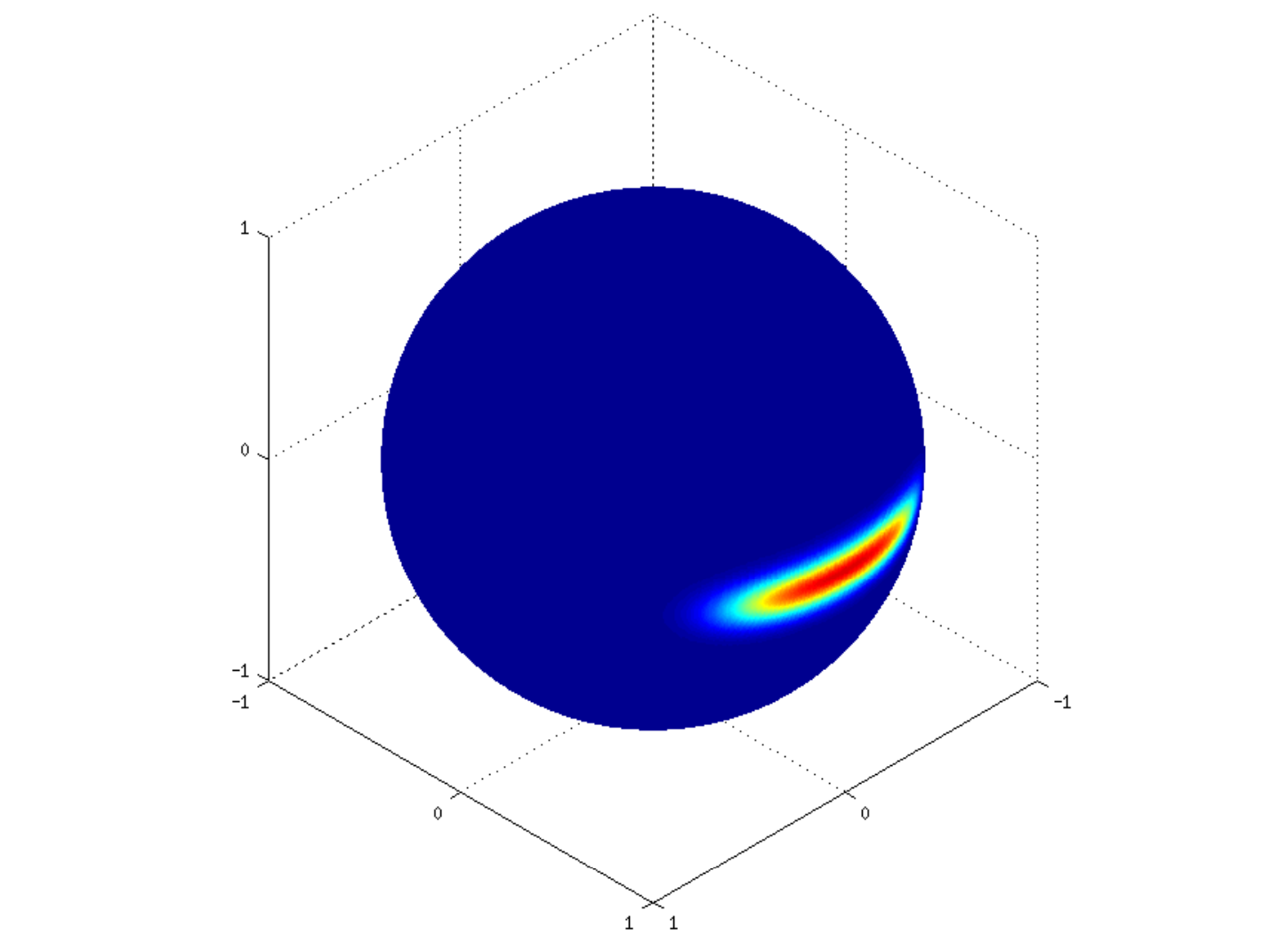}}\hfil
  \hspace{-18mm}
    \subfloat{
        \includegraphics[width=0.1\textwidth]{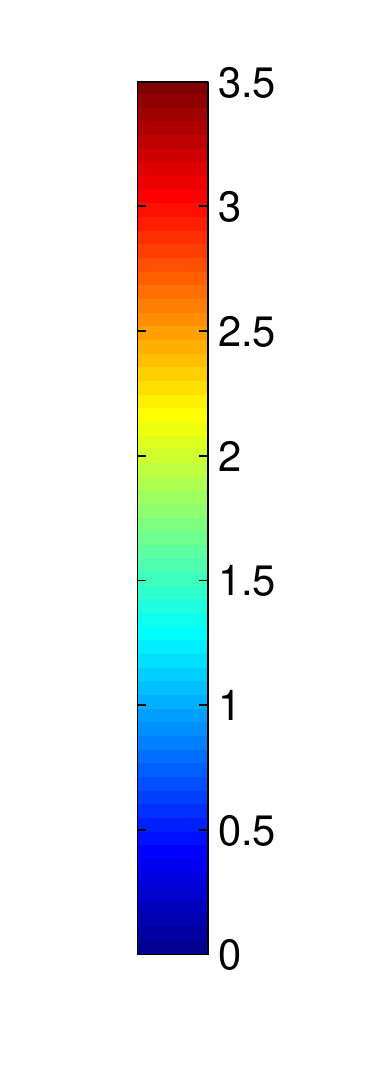}}

    \caption{The Fisher-Bingham five-parameter~(FB5) distribution
$g(\unit{x};\kappa,\unit{\mu},\beta,\A)$, given in \eqref{eq:kent_rotated}, is
plotted on the sphere for parameters $\kappa$, $\beta$, as indicated and: (a)
$\unit{\mu}=[0~0~1]^T$, $\unit{\eta}_1=[0~1~0]^T$, minor axis
$\unit{\eta}_2=[1~0~0]^T$. (b) $\unit{\mu}=[1~0~0]^T$,
$\unit{\eta}_1=[0~1~0]^T$, minor axis $\unit{\eta}_2=[0~0~1]^T$. (c)
$\unit{\mu}=[0~1~0]^T$, $\unit{\eta}_1=[1~0~0]^T$, minor axis
$\unit{\eta}_2=[0~0~1]^T$.}
    \label{fig:FB5}
\end{figure*}

The FB5 distribution is more concentrated and more elliptic for larger values of
$\kappa$ and $\beta$, respectively. As an example, the FB5 distribution
$g(\unit{x};\kappa,\unit{\mu},\beta,\A)$ is plotted on the sphere in
\figref{fig:FB5} for different values of parameters.

The FB5 distribution belongs to the family of spherical distributions in directional statistics and is the analogue of the Euclidean domain bivariate normal distribution with unconstrained covariance matrix~\cite{Kent:1982,Kent:2005}.

\subsection{3D Spatial Fading Correlation~(SFC)}

In MIMO systems, the 3D multipath channel impulse response for a signal arriving
at antenna array is characterized by the steering vector of the antenna array.
For an antenna array consisting of $M$ antenna elements placed at $\bv{z}_p
\in\reals^{3},\,p=1,2,\dotsc,M$, the steering vector, denoted by
$\bv{\alpha}(\unit{x})$, is given by
\begin{equation}
\label{Eq:SV}
\bv{\alpha}(\unit{x})
	= \big[{\alpha}_1(\unit{x}),
{\alpha}_2(\unit{x}),\dotsc{\alpha}_L(\unit{x})\big],\quad
		{\alpha}_p(\unit{x}) \dfn e^{ik \bv{z}_p\cdot\unit{x}},
\end{equation}
where $\unit{x}\in\reals^{3}$ denotes a unit vector pointing in the direction of
wave propagation and $k=2\pi/\lambda$ with $\lambda$ denoting the wavelength of
the arriving signal. For any $h(\unit{x})$ representing the pdf of the angles of
arrival~(AoA) of the multipath components or the unit-normalized power of a signal
received from the direction $\unit{x}$, the 3D SFC function between the $p$-th
and the $q$-th antenna elements, located at $\bv{z}_p$ and $\bv{z}_q$,
respectively, with an assumption that signals arriving at the antenna elements
are narrowband, is given by~\cite{Teal:2002}
\begin{align}\label{eq:spatcorr}
	\rho(\bv{z}_p,\bv{z}_q)
		&\dfn \int_{\untsph}
h(\unit{x})\,\alpha_p(\unit{x})\,\conj{\alpha_q(\unit{x})}\,ds(\unit{x})
\nonumber \\
		&= \int_{\untsph} h(\unit{x})\,e^{ik (\bv{z}_p - \bv{z}_q)
\cdot\unit{x}}\,ds(\unit{x}) \equiv \rho(\bv{z}_p-\bv{z}_q),
\end{align}
which indicates that the SFC only depends on $\bv{z}_p-\bv{z}_q$ and is,
therefore, spatially wide-sense stationary.

\subsection{FB5 Distribution Based Spatial Correlation Model and Problem Under Consideration}

The FB5 distribution offers the flexibility, due to its directional nature, to
model the distribution of normalized power or AoA of the multipath components
for most practical scenarios. Utilizing this capability of FB5 distribution, a
3D spatial correlation model for the mixture of FB5 distributions defining the
AoA distribution has been developed in \cite{Mammasis:2009}. Here the mixture
refers to the positive linear sum of a number of FB5 distributions, each with
different parameters. We defer the formulation of FB5 based correlation model
until~\secref{sec:application}. Although the correlation model proposed in
\cite{Mammasis:2009} is general in a sense that the SFC function can be computed
for any arbitrary antenna geometry, the formulation of SFC function involves the
computation of integrals that can only be carried out using numerical
integration techniques as we highlighted earlier. If the spherical harmonic
expansion of the FB5 distribution is given in closed-form, the SFC function can
be computed analytically following the approach used in \cite{Teal:2002}.

In this paper, we derive an exact expression to compute the spherical harmonic
expansion of the FB5 distribution. Using the spherical harmonic expansion of FB5
distribution, we also formulate exact SFC function for the mixture of FB5
distributions, defining the distribution of AoA of the multipath components. We
also analyse the computational considerations involved in the evaluation of
spherical harmonic expansion of FB5 and distribution and SFC function.


\begin{figure*}[t]
    \centering
    \hspace{-6mm}
    \subfloat[\quad $\kappa=25$, $\beta=10$, \newline
$(\varphi,\vartheta,\omega)=(\pi/2,0,0)$.]{
        \includegraphics[width=0.38\textwidth]{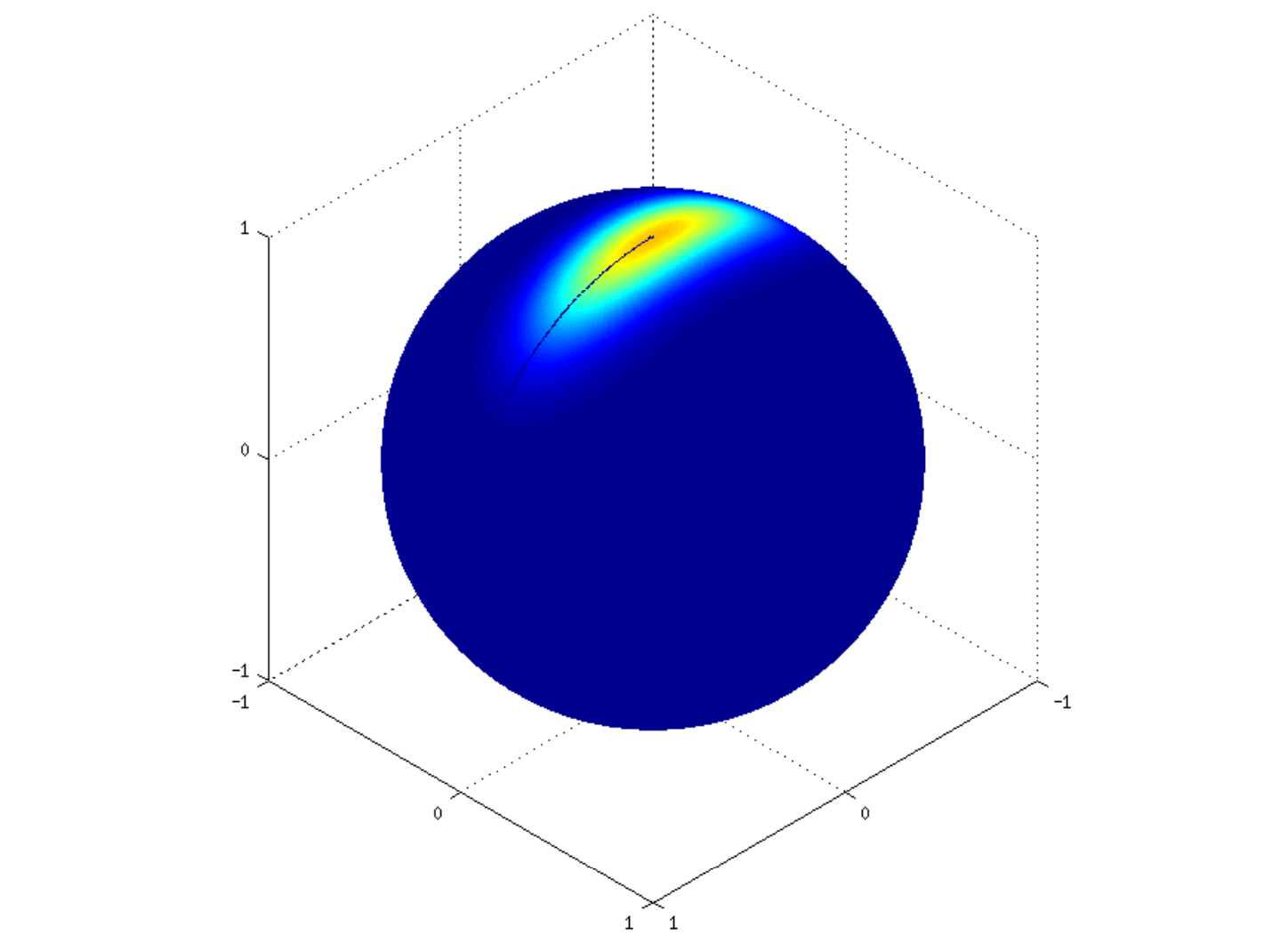}}\hfil
   \hspace{-20mm}
    \subfloat[\quad  $\kappa=100$, $\beta=10$, \newline
$(\varphi,\vartheta,\omega)=(\pi/2,\pi/2,0)$.]{
        \includegraphics[width=0.38\textwidth]{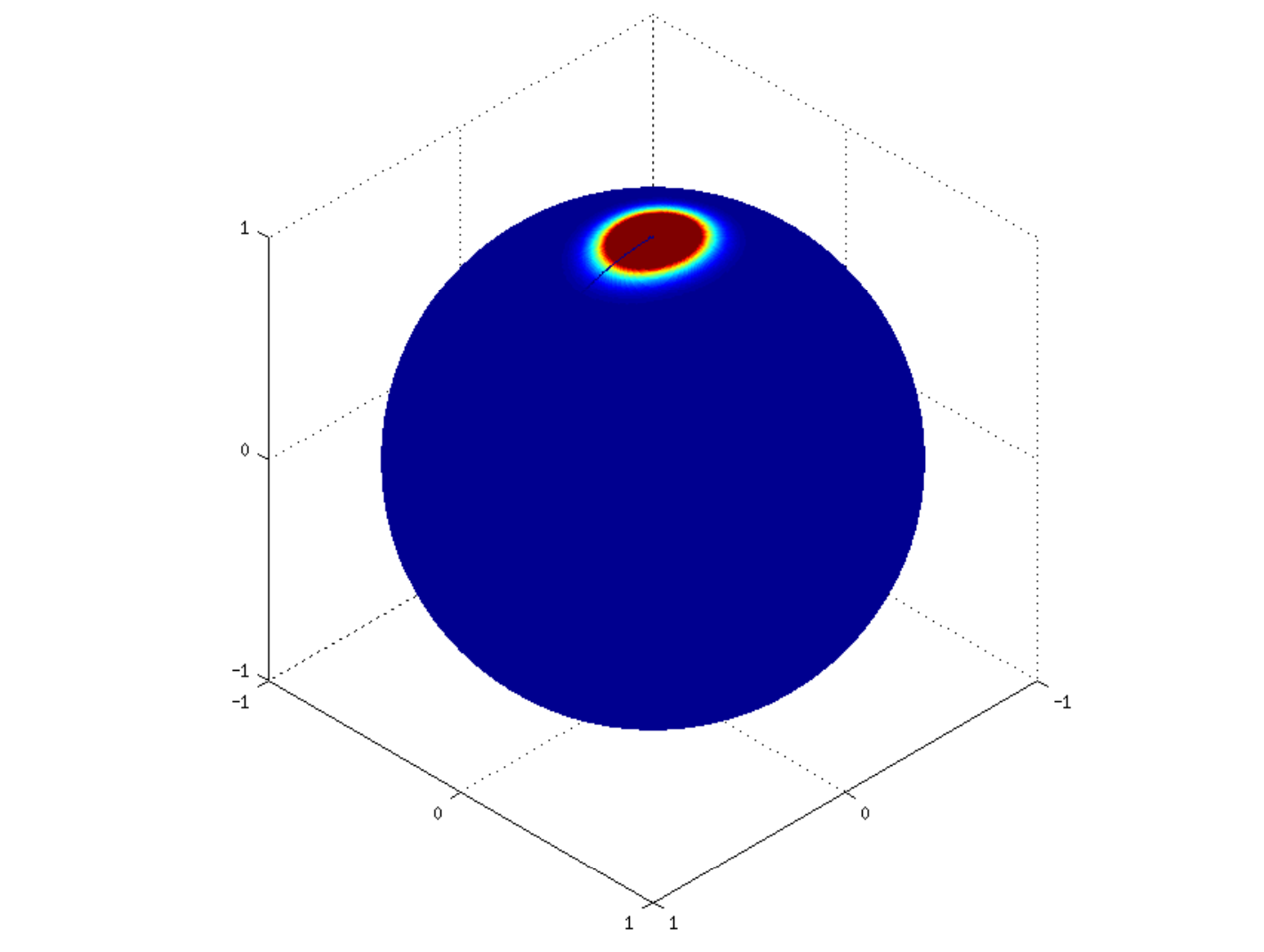}}\hfil
   \hspace{-20mm}
    \subfloat[\quad $\kappa=100$, $\beta=49$, \newline
$(\varphi,\vartheta,\omega)=(\pi/2,\pi/2,\pi/2)$.]{
        \includegraphics[width=0.38\textwidth]{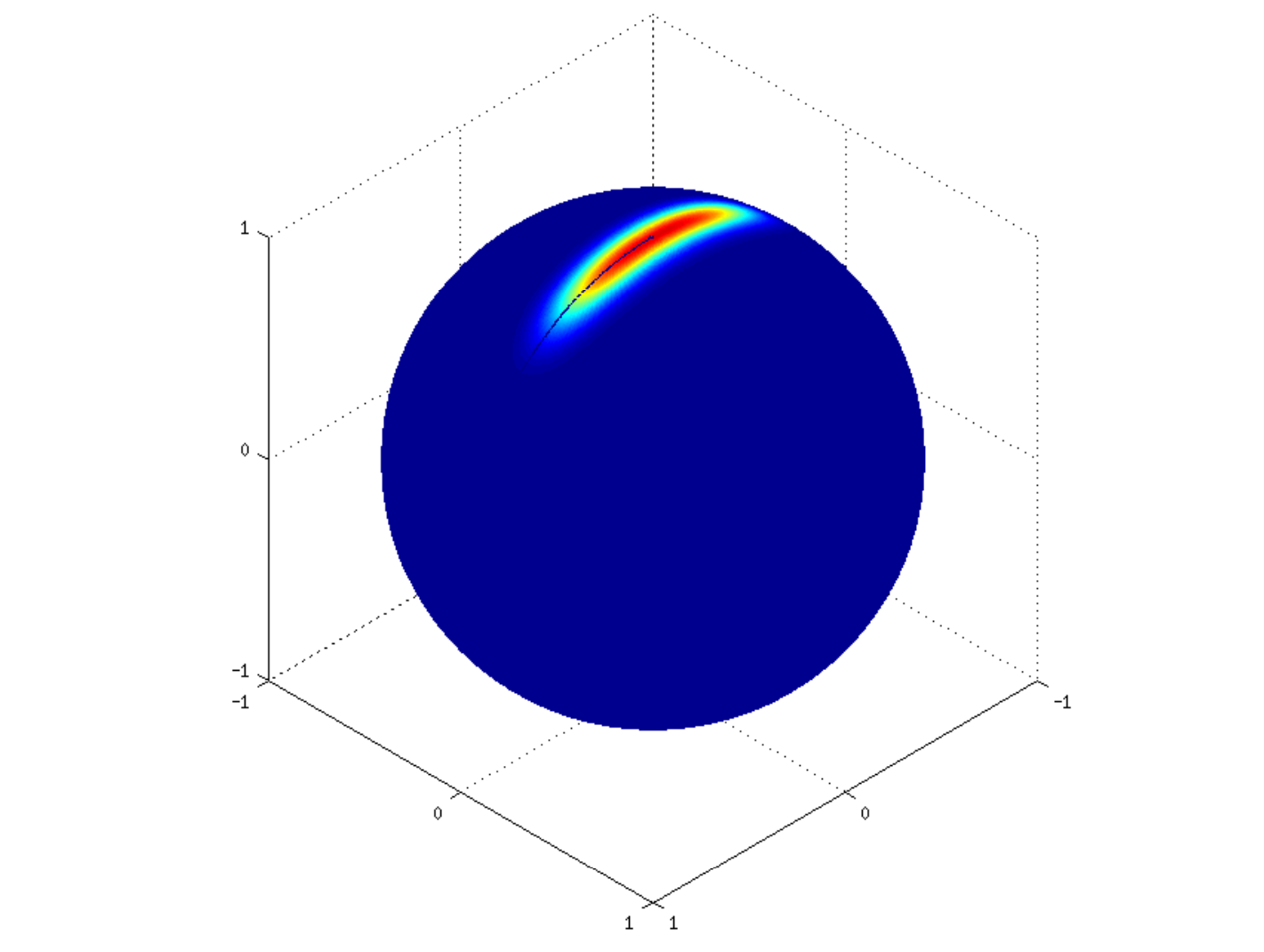}}\hfil
  \hspace{-18mm}
    \subfloat{
        \includegraphics[width=0.1\textwidth]{colorbar}}

    \caption{The standard Fisher-Bingham~(FB) distribution
$f(\unit{x};\kappa,\beta)$, given in \eqref{eq:kent_sph}, is plotted on the
sphere for the concentration parameter $\kappa$ and ovalness parameter $\beta$,
as indicated. Each subfigure is related to the respective subfigure of
\figref{fig:FB5} through the the relation \eqref{eq:kent_rot_op}, with Euler
angles indicated.}
    \label{fig:standardFB}
\end{figure*}

\section{Spherical Harmonic Expansion of the FB5 Distribution}
\label{sec:expansion}

In this section, we derive an analytic expression for the computation of
spherical harmonic coefficients of the FB5 distribution. For convenience, we
determine the spherical harmonic coefficients of the FB5 distribution with
mean~(centre) $\unit{\mu}^0$ located on $z$-axis and major axis
$\unit{\eta}_1^0$ and minor axis $\unit{\eta}_2^0$ aligned along $x$-axis and
$y$-axis respectively, that is,
\begin{equation}\label{Eq:paramters_standard}
  \unit{\mu}^0=[0~0~1]^T\quad
\unit{\eta}_1^0=[1~0~0]^T\quad \unit{\eta}_2^0=[0~1~0]^T.
\end{equation}
With the centre, major and minor axes as given in \eqref{Eq:paramters_standard},
the FB5 distribution, referred to as the \emph{standard} Fisher-Bingham~(FB)
distribution, has the pdf given by
\begin{equation}\label{eq:kent_sph}
f(\unit{x};\kappa,\beta) \dfn \frac{1} { C(\kappa , \beta)} e^{\kappa\cos\theta +
\beta\sin^2\theta\cos 2\phi},
\end{equation}
which is related to the FB5 distribution pdf
$g(\unit{x};\kappa,\beta,\unit{\mu},\A)$ given in
\eqref{eq:kent_rotated} through the rotation operator
$\mathcal{D}(\varphi,\vartheta,\omega)$ as
\begin{equation}\label{eq:kent_rot_op}
 g(\unit{x};\kappa,\beta,\unit{\mu},\A)=\left(\mathcal{D}(\varphi,\vartheta,\omega)f\right) (\unit{x
}; \kappa,\beta) = f(\bv{R}^{-1}\unit{x}; \kappa,\beta),
\end{equation}
where the rotation matrix is related to $\unit{\mu},~\unit{\eta}_1,~\unit{\eta}_2$~(parameters of FB5 distribution) as
\begin{equation}\label{Eq:Rmat_FB_parameters}
\bv{R}=[\unit{\eta}_1 , \unit{\eta}_2, \unit{\mu}],
\end{equation}
and the Euler angles $(\varphi,\vartheta,\omega)$ are related to the rotation matrix $\bv{R}$ through \eqref{eq:rot_matrix}.
As an example, we rotate each of the FB5 distribution
$g(\unit{x};\kappa,\unit{\mu},\beta,\A)$ plotted in \figref{fig:FB5} to obtain
the standard Fisher-Bingham distribution $f(\unit{x};\kappa,\beta)$, plotted in
\figref{fig:standardFB}, where we have indicated the Euler angles that relate
the FB5 distribution and the Fisher-Bingham distribution
through~\eqref{eq:kent_rot_op}.

\subsection{Spherical Harmonic Expansion of Standard FB distribution}

Here, we derive a closed-form expression to compute the spherical harmonic
coefficients of the standard FB distribution, given in \eqref{eq:kent_sph}.
Later in this section, noting the relation between standard FB distribution and
FB5 distribution given in \eqref{eq:kent_rot_op}, and the effect of the rotation
operation on the spherical harmonic coefficients, we determine the coefficients
of FB5 distribution.

The spherical harmonic coefficient, denoted by $\shc{f}{\ell}{m}$, of the
standard Fisher-Bingham distribution given in \eqref{eq:kent_sph} can be
expressed as
\begin{equation}\label{eq:projection}
\shc{f}{\ell}{m} \dfn \innerp{f(\,\cdot\,; \kappa,\beta)}{Y_\ell^m} =  \intsph f(\unit{x
}; \kappa,\beta) \overline{Y_\ell^m(\unit{x})}
ds(\unit{x}).
\end{equation}
Since $f(\unit{x};\kappa,\beta)$ is a real function, we only need to compute the
spherical harmonic coefficients for positive orders $0 \le m\le \ell$ for each
degree $\ell\ge 0$. The coefficients for the negative orders can be readily
computed using the conjugate symmetry relation, noted in
\eqref{Eq:conj_symmetry}.

To derive a closed-form expression for computing the spherical harmonic
coefficients, we rewrite \eqref{eq:projection} explicitly as
\ifCLASSOPTIONonecolumn
\begin{equation}
 \begin{split}
\shc{f}{\ell}{m} &= \frac{N_{\ell}^{m}}{C(\kappa,\beta)}  \int_{0}^{\pi}
e^{\kappa\cos\theta}
P_{\ell}^{m}(\cos\theta)\,  \sin\theta d\theta \int_{0}^{2\pi}
e^{-im\phi} e^{\beta \sin^2 \theta \cos 2\phi} d\phi\\
                 &= \frac{N_{\ell}^{m}}{C(\kappa,\beta)} \int_{0}^{\pi}
e^{\kappa \cos\theta} \,F_m(\theta)\,P_{\ell}^{m}(\cos\theta) \sin\theta
d\theta,
\end{split}
\end{equation}
\else
\begin{equation}
 \begin{split}
\shc{f}{\ell}{m} &= \frac{N_{\ell}^{m}}{C(\kappa,\beta)}  \int_{0}^{\pi}
e^{\kappa\cos\theta}
P_{\ell}^{m}(\cos\theta)\,  \sin\theta d\theta  \nonumber \\ &\quad \quad\times
\int_{0}^{2\pi} e^{-im\phi} e^{\beta \sin^2 \theta \cos 2\phi} d\phi\\
                 &= \frac{N_{\ell}^{m}}{C(\kappa,\beta)} \int_{0}^{\pi}
e^{\kappa \cos\theta} \,F_m(\theta)\,P_{\ell}^{m}(\cos\theta) \sin\theta
d\theta,
\end{split}
\end{equation}
\fi

\noindent where
\begin{equation*}\label{eq:integralPhi}
F_m(\theta) = \int_{0}^{2\pi}e^{\beta \sin^2 \theta \cos 2\phi - im\phi}
d\phi,
\end{equation*}

\noindent which we evaluate as~\footnote{Using Mathematica.}
\begin{equation*}
\begin{split}
F_m(\theta) &=  \begin{cases}
                2\pi I_{m/2}(\beta\sin^2\theta) & m \in {0,\,2,\,4,\,\hdots} \\
                0 & m \in {1,\,3,\,5,\,\hdots}.
                \end{cases}
\end{split}
\end{equation*}
By expanding the modified Bessel function $I_{m/2}(\beta\sin^2\theta)$ as
\begin{equation}\label{eq:Bessel_expansion}
I_{m/2}(\beta\sin^2\theta) = \sum_{t=0}^\infty
\left(\frac{\beta}{2}\right)^{2t+m/2}
\frac{(\sin\theta)^{4t+m}}{t!(t+m/2)!},
\end{equation}

\noindent the exponential $e^{\kappa\cos\theta}$ as \cite{Mammasis:2008}
\begin{equation}\label{eq:kappa_expansion}
e^{\kappa\cos\theta} = \sqrt{\frac{\pi}{2\kappa}}\sum_{n=0}^\infty (2n+1)
I_{n+1/2}(\kappa) P_{n}^{0}(\cos\theta),
\end{equation}
and using the relation between associated Legendre polynomial and Wigner-$d$
function given in \eqref{Eq:Plm_Wignerd}, along with the following expansion of
Wigner-$d$ function in terms of complex exponentials
\begin{equation}
\label{eq:Wignerd-risbo}
	d^{\ell}_{m,n}(\theta) = i^{n-m} \sum_{u=-\ell}^{\ell}
	d_{u,m}^{\ell}({\pi}/{2})\,d_{u,n}^{\ell}({\pi}/{2})\,e^{iu\theta},
\end{equation}
we obtain a closed-form
expression for the spherical harmonic coefficient in \eqref{eq:projection} as
\ifCLASSOPTIONonecolumn
\begin{equation}\label{eq:coeff}
\begin{split}
\shc{f}{\ell}{m} &=
\frac{\pi i^{-m}}{C(\kappa,\beta)}\sqrt{\frac{2\ell+1}{2\kappa}}
\sum_{n=0}^{\infty} (2n+1)I_{n+1/2}(\kappa)
\sum_{u=-n}^{n}  \left(d_{u,0}^{n}({\pi}/{2})\right)^2
\sum_{t=0}^{\infty}\frac{(\beta/2)^{2t+m/2}}{ \Gamma(t+1)\Gamma(t+m/2+1)}
~\times\\
& \hspace{70mm}\sum_{u'=-\ell}^{\ell} d_{u',0}^{\ell}({\pi}/{2})
d_{u',m}^{\ell}({\pi}/{2})\, G(4t+m+1,u+u'),
\end{split}
\end{equation}
\else
\begin{equation}\label{eq:coeff}
\begin{split}
\shc{f}{\ell}{m} &=
\frac{\pi i^{-m}}{C(\kappa,\beta)}\sqrt{\frac{2\ell+1}{2\kappa}}
\sum_{n=0}^{\infty} (2n+1)I_{n+1/2}(\kappa)~\times\\
&\sum_{u=-n}^{n}  \left(d_{u,0}^{n}({\pi}/{2})\right)^2
\sum_{t=0}^{\infty}\frac{(\beta/2)^{2t+m/2}}{ \Gamma(t+1)\Gamma(t+m/2+1)}
~\times\\
& \sum_{u'=-\ell}^{\ell} d_{u',0}^{\ell}({\pi}/{2})
d_{u',m}^{\ell}({\pi}/{2})\, G(4t+m+1,u+u'),
\end{split}
\end{equation}
\fi
\noindent where $\Gamma(\cdot)$ denotes the Gamma function and we have used the
following identity~\cite[Sec.\,3.892]{Jeffrey:2007}
\begin{align}\label{Eq:intergal_theta}
G(p,q) &\dfn \int_{0}^{\pi} (\sin\theta)^p e^{i q \theta} d\theta, \nonumber \\
       & = \frac{\pi e^{iq\pi/2}\Gamma(p+2)}
	   {2^{p}(p+1)\Gamma(\frac{p+q+2}{2})\Gamma(\frac{p-q+2}{2})}\,.
\end{align}

\subsection{Spherical Harmonic Expansion of FB5 distribution}

We use the relation between standard FB distribution and FB5 distribution given in \eqref{eq:kent_rot_op} to determine the spherical harmonic expansion of FB5 distribution. The Euler angles $(\varphi,\vartheta,\omega)$ in \eqref{eq:kent_rot_op}, which characterize the relation between the standard FB distribution and the FB5 distribution, can be obtained from the rotation matrix $\bv{R}$ formulated in \eqref{Eq:Rmat_FB_parameters} in terms of the parameters of FB5 distribution. By comparing \eqref{eq:rot_matrix} and \eqref{Eq:Rmat_FB_parameters}, $\vartheta $ is given by
\begin{equation}\label{Eq:extraction_angle_1}
 \vartheta=\cos^{-1}(R_{3,3}),
\end{equation}
where $R_{a,b}$ denotes the entry at $a$-th row and $b$-th column of the matrix $\bv{R}$ given in \eqref{Eq:Rmat_FB_parameters}.
Similarly $\varphi\in[0,2\pi)$ and $\omega\in[0,2\pi)$ can be found by a four-quadrant
search satisfying
\begin{equation}\label{Eq:extraction_angle_2}
  \sin\varphi =\frac{R_{2,3}}{\sqrt{1-(R_{3,3})^2}},\quad
  \cos\varphi =\frac{R_{1,3}}{\sqrt{1-(R_{3,3})^2}},
\end{equation}
\begin{equation}\label{Eq:extraction_angle_3}
  \sin\omega =\frac{R_{3,2}}{\sqrt{1-(R_{3,3})^2}},\quad
  \cos\omega =\frac{-R_{3,1}}{\sqrt{1-(R_{3,3})^2}},
\end{equation}
respectively.

Once the Euler angles $(\varphi,\vartheta,\omega)$ are extracted from
the parameters $\unit{\mu},\unit{\eta}_1, \unit{\eta}_2$ using \eqref{Eq:extraction_angle_1}--\eqref{Eq:extraction_angle_3}, the spherical harmonic
coefficients of the FB5 distribution $g(\unit{x};\unit{\mu},\kappa,\A)$ given in \eqref{eq:kent_rotated} can be
computed following the effect of the rotation operation on the spherical harmonic coefficients as~\cite{Kennedy-book:2013}
\begin{align}\label{eq:kent_rotated_shc}
\shc{g}{\ell}{m}&\dfn  \innerp{g(\,\cdot\,;\unit{\mu},\kappa,\A)}{Y_\ell^m} =
\innerp{\mathcal{D}(\varphi,\vartheta,\omega)f}{Y_\ell^m} \nonumber \\
&= \summp  D^{\ell}_{m,m'} (\varphi,\vartheta,\omega)\shc{f}{\ell}{m'},
\end{align}
where $\shc{f}{\ell}{m'}$ is the spherical harmonic coefficient of degree
$\ell$ and order $m'$ of the standard Fisher-Bingham distribution $f(\unit{x})$.
In \eqref{eq:kent_rotated_shc},
$ D^{\ell}_{m,m'}(\varphi,\vartheta,\omega)$ denotes the Wigner-$D$
function of degree $\ell$ and orders $|m|,|m'|\le\ell$ and
is given by~\cite{Kennedy-book:2013}
\begin{equation}\label{eq:rot_hram}
 D^{\ell}_{m,m'}(\varphi,\vartheta,\omega)=\mathrm{e}^{-im\varphi}d^{\ell}_{m, m'}
(\vartheta)\mathrm{e}^{-im'\omega}.
\end{equation}

\subsection{Computational Considerations}

Here we discuss the computation of Wigner-$d$ functions, at a fixed argument of
$\pi/2$, which are essentially required for the computation of spherical
harmonic expansion of standard FB or FB5 distribution using the proposed
formulation~\eqref{eq:coeff}. Another computational consideration that is
addressed here is the evaluation of infinite summations over $t$ and $n$
involved in the computation of \eqref{eq:coeff}.

\subsubsection{Computation of Wigner-$d$ functions}
For the computation of spherical harmonic coefficients of the standard FB
distribution using \eqref{eq:coeff}, we are required to compute the Wigner-$d$
functions at a fixed argument of $\pi/2$, that is $d^{\ell}_{u,m}(\pi/2)$, for each
$|u|,\,|m|\le\ell$. Let $\mathbf{D}_{\ell}$
denote the matrix of size $(2\ell+1)\times(2\ell+1)$ with entries
$d^{\ell}_{u,m}(\pi/2)$ for $|u|,|m|\le \ell$. The matrix $\mathbf{D}_{\ell}$
can be computed for each $\ell=1,2,\dotsc,$ using the relation given
in~\cite{Trapani:2008} that recursively computes $\mathbf{D}_{\ell}$ from $\mathbf{D}_{\ell-1}$.

\subsubsection{Truncation over $n$}
\begin{figure}[tbp]
\centering
\includegraphics[width=0.85\columnwidth]{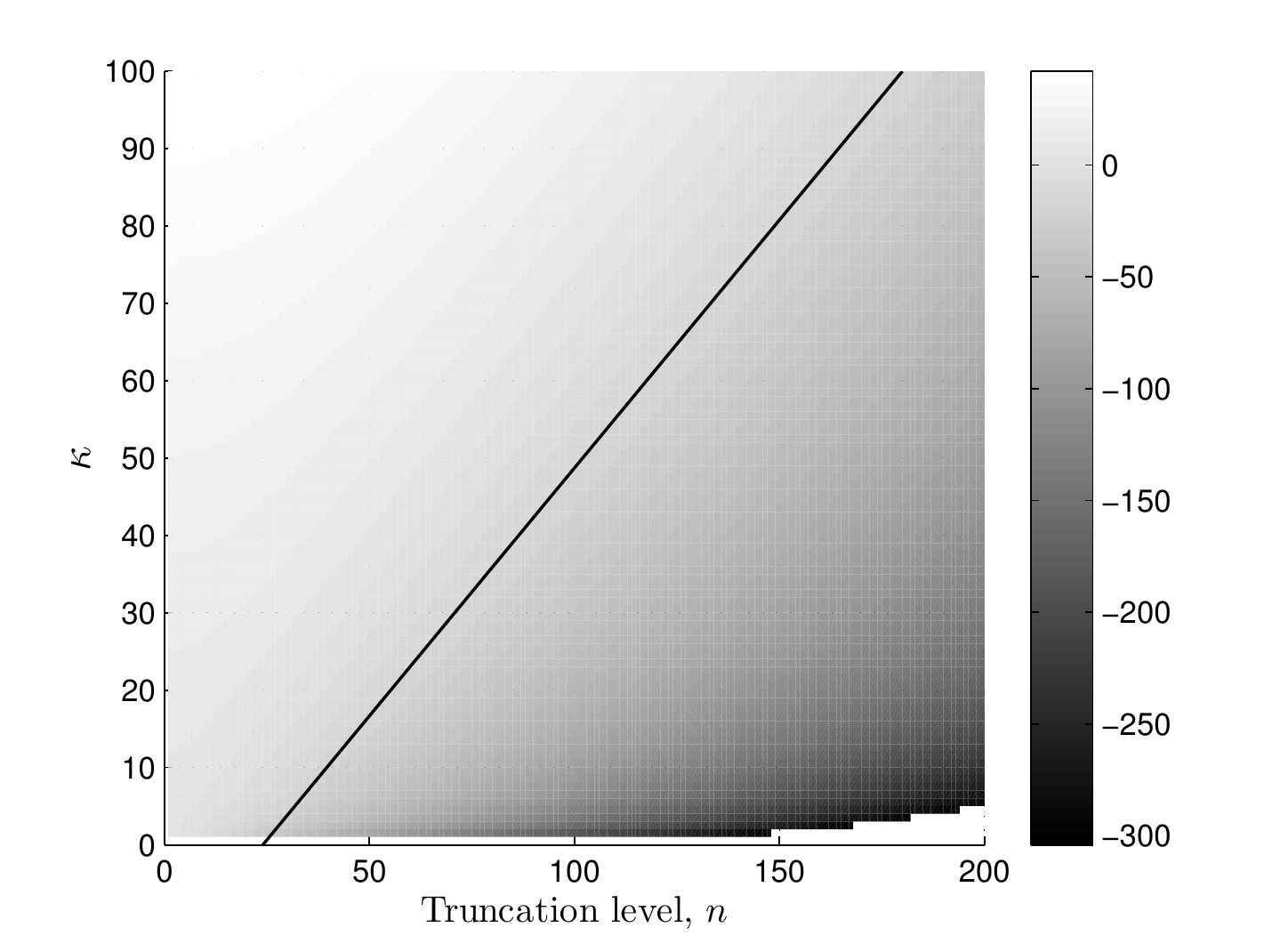}
\caption{Surface plot of modified Bessel function $I_{n+1/2}(\kappa)$
for different values of $\kappa$ and $n$. The region $I_{n+1/2}(\kappa)<10^{-16}$ is approximated by the region
below the straight line, $n=\frac{3}{2}\kappa + 24$. Only those values
$I_{n+1/2}(\kappa)>10^{-16}$ plotted above the straight line account for the
sum in \eqref{eq:kappa_expansion} to make it accurate to the machine precision
level.}
\label{fig:n_surface}
\end{figure}

The infinite sum over $n$ arises from \eqref{eq:kappa_expansion}
where an exponential function of the concentration measure $\kappa$ and the co-latitude
angle $\theta$ is expanded as an infinite sum of the modified Bessel functions
and Legendre polynomials. It is well known that the Legendre polynomials are
oscillating functions with a maximum value of one. The modified Bessel function
$I_{n+1/2}(\kappa)$
that makes up the core of the expansion given in \eqref{eq:kappa_expansion}
decays quickly~(and \emph{monotonically}) to zero as $n\rightarrow\infty$ for a
given $\kappa$. We use this feature of the modified Bessel function
$I_{n+1/2}(\kappa)$ to truncate the summation over $n$. We plot
$I_{n+1/2}(\kappa)$ for different values of $n$ and  $0\le \kappa\le 100$ in
\figref{fig:n_surface}, where it is evident that the Bessel function quickly
decays to zero. We propose to truncate the summation over $n$ in
\eqref{eq:kappa_expansion}, or equivalently in \eqref{eq:coeff}, at $n=N$ such
that $I_{N+1/2}(\kappa)<10^{-16}$~(double machine precision). We
approximate the linear relationship between such truncation level and the
concentration parameter $\kappa$, given by
\begin{equation}\label{eq:n_truncation}
 N=\frac{3}{2}\kappa + 24,
\end{equation}
which is also indicated in \figref{fig:n_surface}. This truncation level is found to give truncation error less than the machine precision level for the values of concentration parameter $\kappa$ in the range $0 \le \kappa \le 100$~(used in practice~\cite{Mammasis:2008,Mammasis:2009}). We note that the truncation error becomes smaller for large values of $\kappa$, indicating that the truncation at level less than $N$ given in \eqref{eq:n_truncation} may also allow sufficiently accurate computation of spherical harmonic coefficients. Further analysis on establishing the relationship between the concentration parameter $\kappa$ and the truncation level is beyond the scope of current work.

\subsubsection{Truncation over $t$}
\begin{figure}[tbp]
\centering
\includegraphics[width=0.85\columnwidth]{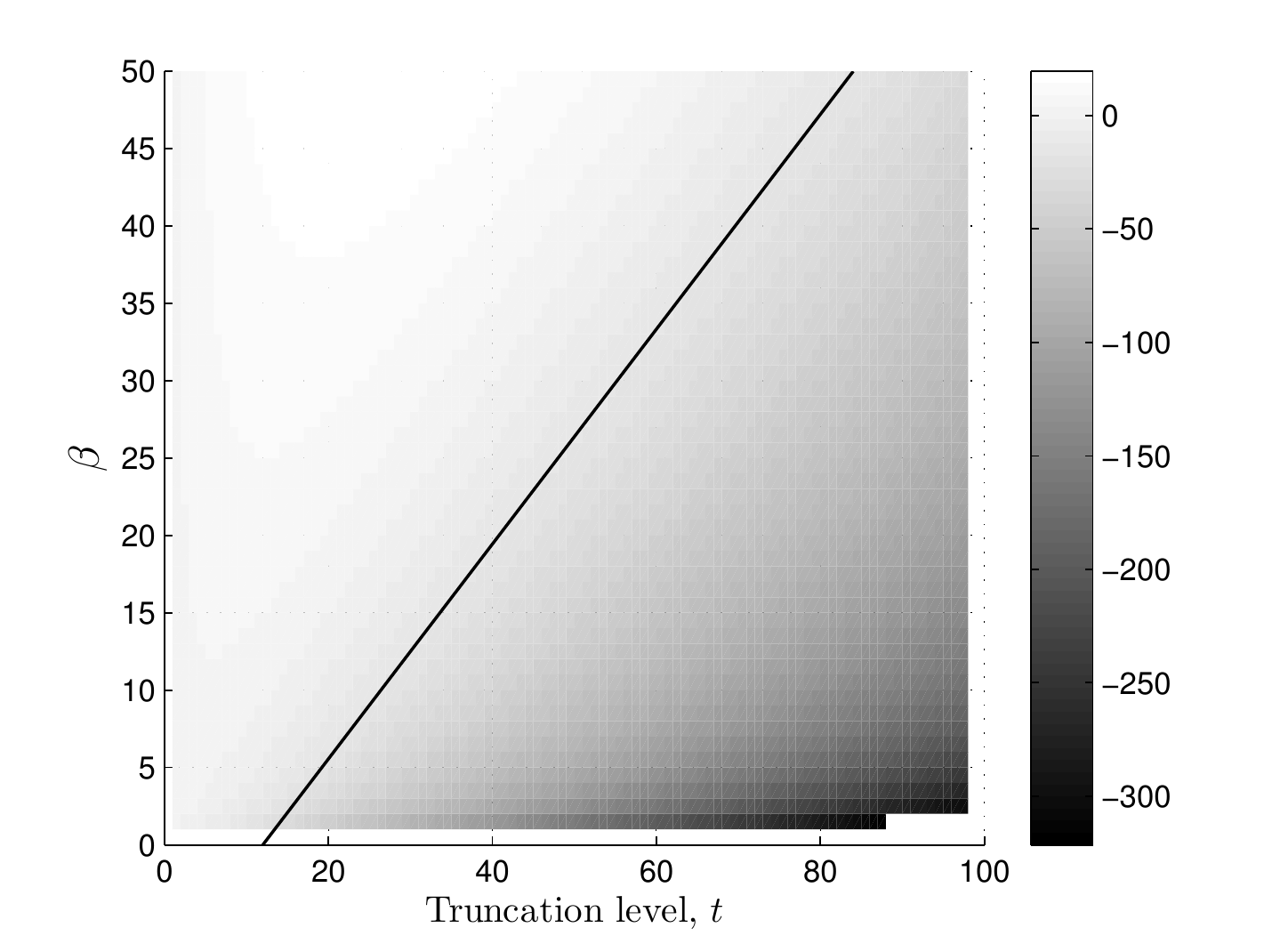}
\caption{
Surface plot of the term inside the summation on the right hand side of
\eqref{eq:Bessel_expansion} for $m=0$ and $\theta=\pi/2$, that is,
$S(\beta,t)\dfn\frac{\beta^{2t}}{t!t!\, 2^{2t}}$ for different values of the
ovalness parameter $\beta$ and $t$. The region $R(\beta,t)<10^{-16}$ is
approximated by the region below the straight line, $t = \frac{36}{25}\beta +
12$.}
\label{fig:t_surface}
\end{figure}

The expansion of modified Bessel functions of the first kind for a given value
of $\beta$ as given in \eqref{eq:Bessel_expansion} introduces the infinite sum
over $t$. Again, the decaying~(non-monotonic in this case) characteristics of
the term inside the summation on right hand side of \eqref{eq:Bessel_expansion} with the increase in $t$ makes it possible to truncate the infinite sum given in \eqref{eq:Bessel_expansion} that mainly depends on the ovalness parameter
$\beta$ with a minimal error. We plot the term inside the summation on the right hand side of
\eqref{eq:Bessel_expansion} for $m=0$ and $\theta=\pi/2$, that is,
$S(\beta,t)\dfn\frac{\beta^{2t}}{t!t!\, 2^{2t}}$ for different values of
the ovalness parameter $\beta$ in \figref{fig:t_surface}. We again propose to
truncate the summation over $t$ at the truncation level $T$ such that the
terms after the truncation level $T$ are less than $10^{-16}$ and do not
have impact on the summation. We approximate the linear relationship between the
ovalness parameter $\beta$ and the truncation level $T$

\begin{equation}\label{eq:t_truncation}
 T=\frac{36}{25}\beta + 12.
\end{equation}
Since we are computing coefficients for positive orders $0 \le m\le \ell$ for each degree $\ell$, the truncation level given by \eqref{eq:t_truncation}, that is obtained for order $m=0$ and $\theta=\pi/2$, for the summation in \eqref{eq:Bessel_expansion}, is also valid for all positive orders $0 \le m\le \ell$ and all $\theta\in[0,\pi]$ because the term inside the summation is maximum for $m=0$ and $\theta=\pi/2$ for a given summation variable $t$.

We finally note that the truncation levels proposed here allow accurate computation of spherical harmonic coefficients to the level of double machine precision. Later in the paper, we validate the correctness of the derived expression for the computation of spherical harmonic coefficients of the standard FB distribution.


\section{3D Spatial Fading Correlation for Mixture of Fisher-Bingham Distributions}
\label{sec:application}

In this section, we derive a closed form expression to compute the SFC function
for the spatial correlation model based on a mixture of FB5 distributions
defining the distribution of AoA of multipath
components~\cite{Mammasis:2008,Mammasis:2009}. Let $h(\unit{x})$ denotes the pdf
of the distribution defined as a positive linear sum of $W$ FB5 distributions,
each with different parameters, and is given by
\begin{align}\label{Eq:Mixture_pdf}
h(\unit{x}) = \sum_{w=1}^{W} K_w  g(\unit{x};\kappa_w,\unit{\mu}_w,\beta_w,\A_w),
\end{align}
where $K_w$ denotes the weight of $w$-th FB5 distribution with parameters
$\kappa_w,\unit{\mu}_w,\beta_w$ and $\A_w$. We assume that each $K_w$ is
normalized such that $\|h\|_2=1$. For the distribution with pdf defined in
\eqref{Eq:Mixture_pdf}, the SFC function, given in \eqref{eq:spatcorr} has been
formulated in \cite{Mammasis:2009} and analysed for a uniform circular antenna
array but the integrals involved in the SFC function were numerically
computed.

We follow the approach introduced in \cite{Teal:2002} to derive a closed-form
3D SFC  function using the proposed closed-form spherical harmonic expansion of
the FB5 distribution. Using spherical harmonic expansion of plane
waves~\cite{Colton:2013}:

\begin{equation*}\label{eq:wave_exp}
	e^{ik\bv{z}_p\cdot\unit{x}}=4\pi\sum_{\ell=0}^{\infty}i^{\ell}j_{\ell}
\parn[\big]{k\norm{\bv{z}_p}} \summ
Y_{\ell}^{m}\parn[\big]{\bv{z}_p/\norm{\bv{z}_p}}
\conj{Y_{\ell}^{m}(\unit{x})},
\end{equation*}
and expanding the mixture distribution $h(\unit{x})$ in \eqref{Eq:Mixture_pdf},
following \eqref{eq:f_expansion}, and employing the orthonormality of spherical
harmonics, we write the SFC function in \eqref{eq:spatcorr} as~\cite{Teal:2002}
\ifCLASSOPTIONonecolumn
\begin{align}\label{Eq:sfc_analytic}
 \rho(\bv{z}_p-\bv{z}_q)= 4\pi \sum_{\ell=0}^{\infty}i^{\ell}	
j_{\ell}\parn[\big]{k\norm{\bm{z}_{p}-\bm{z}_{q}}} \summ \shc{h}{\ell}{m}
Y_{\ell}^{m}\parn[\Big]{\frac{\bm{z}_{p}-\bm{z}_{q}}{\norm{\bm{z}_{p}-\bm{z}_{q}
}}},
\end{align}
where
\begin{align}\label{Eq:shc_mixture}
\shc{h}{\ell}{m} &= \innerp{h}{Y_\ell^m} \nonumber \\
&= \sum_{w=1}^{W}~\summp  K_w  D^{\ell}_{m,m'}\,
(\varphi_w,\vartheta_w,\omega_w)\shc{f}{\ell}{m'},
\end{align}
\else
\begin{align}\label{Eq:sfc_analytic}
 \rho(\bv{z}_p-\bv{z}_q)		&= 4\pi \sum_{\ell=0}^{\infty}i^{\ell}		j_{\ell}\parn[\big]{k\norm{\bm{z}_{p}-\bm{z}_{q}}}~\times \nonumber \\
&\quad \summ \shc{h}{\ell}{m}
Y_{\ell}^{m}\parn[\Big]{\frac{\bm{z}_{p}-\bm{z}_{q}}{\norm{\bm{z}_{p}-\bm{z}_{q}
}}},
\end{align}
where
\begin{align}\label{Eq:shc_mixture}
\shc{h}{\ell}{m} &= \innerp{h}{Y_\ell^m} \nonumber \\
&= \sum_{w=1}^{W}\quad \summp  K_w  D^{\ell}_{m,m'}\, (\varphi_w,\vartheta_w,\omega_w)\shc{f}{\ell}{m'},
\end{align}
\fi
which is obtained by combining \eqref{eq:kent_rotated_shc} and
\eqref{Eq:Mixture_pdf}. Here, $\shc{f}{\ell}{m'}$ denotes the spherical harmonic
coefficient of the standard FB distribution and the Euler angles
$(\varphi_w,\vartheta_w,\omega_w)$ relate the $w$-th FB5 distribution
$g(\unit{x};\kappa_w,\unit{\mu}_w,\beta_w,\A_w)$ of the mixture and the standard
FB distribution $f(\unit{x})$ through \eqref{eq:kent_rot_op}. In the computation
of the SFC using the proposed formulation, given in \eqref{Eq:sfc_analytic}, we
note that the summation for $\ell$ over first few terms yields sufficient
accuracy as higher order Bessel functions decay rapidly to zero for points near
each other in space, as indicated in~\cite{Teal:2002,Jones:2002}. We conclude
this section with a note that the SFC function can be analytically computed
using the proposed formulation for an arbitrary antenna array geometry and the
distribution of AoA modelled by a mixture of FB5 distributions, each with
different parameters. In the next section, we evaluate the proposed SFC function
for uniform circular array and a 3D regular dodecahedron array of antenna
elements.

\section{Experimental Analysis}\label{sec:analysis}

We conduct numerical experiments to validate the correctness of the proposed
analytic expressions, formulated in \eqref{eq:coeff}--\eqref{Eq:intergal_theta}
and \eqref{Eq:sfc_analytic}--\eqref{Eq:shc_mixture}, for the computation of the
spherical harmonic coefficients of standard FB distribution and the SFC function
for a mixture of FB5 distributions defining the distribution of AoA,
respectively. For computing the spherical harmonic transform and discretization
on the sphere, we employ the recently developed optimal-dimensionality sampling
scheme on the sphere~\cite{Khalid:2014}. Our \text{\matlab} based code to
compute the spherical harmonic coefficients of the standard FB~(or FB5)
distribution and the SFC function using the results and/or formulations
presented in this paper, is made publicly available.

\begin{figure}[tbp]
\centering
\includegraphics[width=0.75\columnwidth]{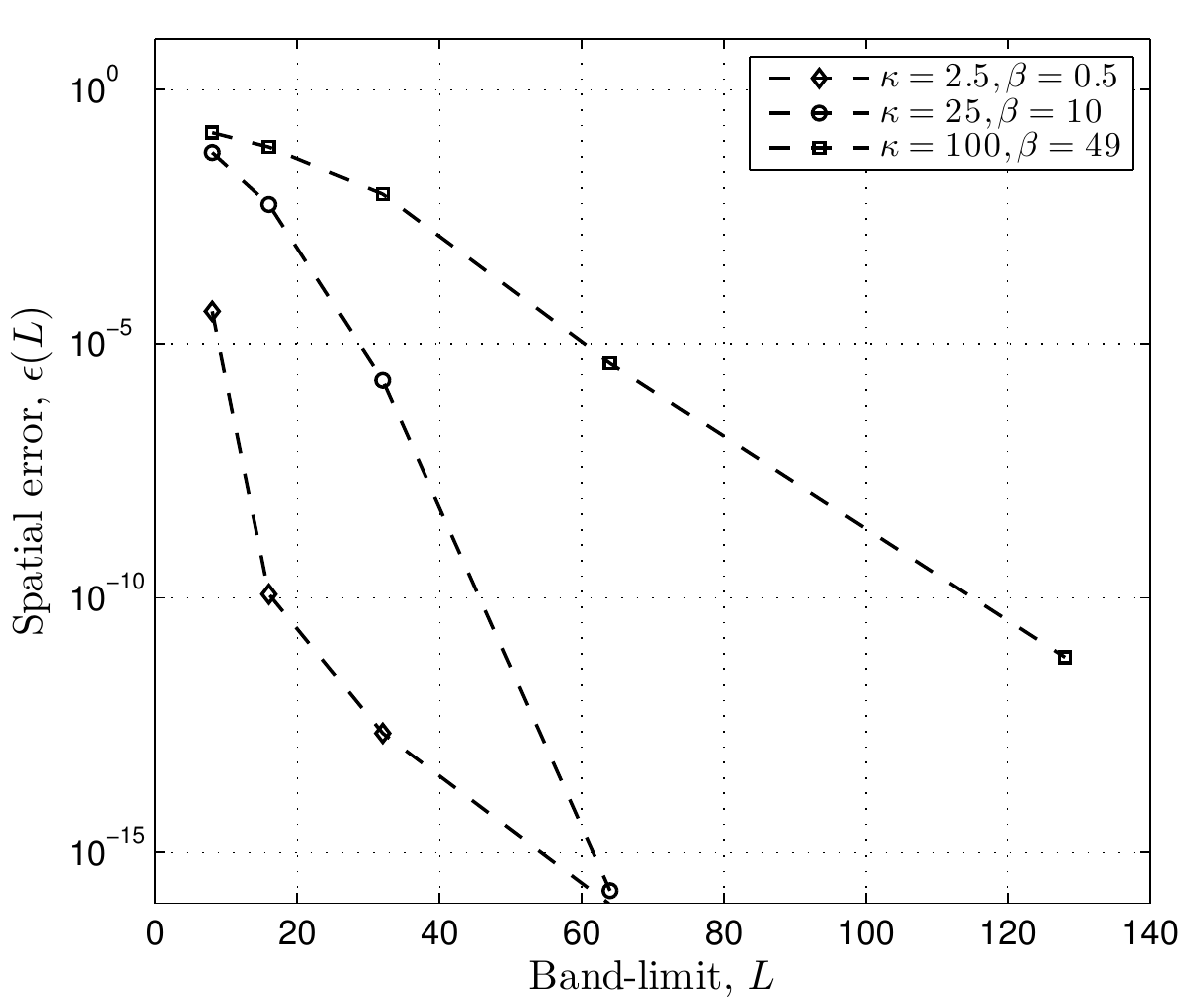}
\caption{Spatial error $\epsilon(L)$, given in \eqref{eq:spatial_error}, between the standard FB distribution given in \eqref{eq:kent_sph} and the reconstructed standard FB distribution from its coefficients up to the degree $L-1$. The convergence of the spatial error to zero~(machine precision), as $L$ increases, corroborates the correctness of the derived spherical harmonic expansion of standard FB distribution. }
\label{fig:spatial_error}
\end{figure}

\subsection{Accuracy Analysis - Spherical Harmonic Expansion of standard FB distribution}
In order to analyse the accuracy of the proposed analytical expression, for the
computation of spherical harmonic coefficients of the standard FB distribution,
given in \eqref{eq:coeff}, we define the spatial error as
\begin{equation}\label{eq:spatial_error}
 \epsilon(L) =
\frac{1}{L^2}\sum_{\unit{x}_p}\big| f(\unit{x}_p)- \sum_{\ell=0}^{L-1} \sum_{m=-\ell}^{\ell} \shc{f}{\ell}{m} Y_\ell^m(\unit{x}_p) \big|^2 ,
\end{equation}
which quantifies the error between the standard FB distribution given in \eqref{eq:kent_sph} and the reconstructed standard FB distribution from its coefficients, computed using the proposed analytic expression \eqref{eq:coeff}, up to the degree $L-1$. The summation in \eqref{eq:spatial_error} is averaged over $L^2$ number of samples of the sampling scheme~\cite{Khalid:2014}. We plot the spatial error $ \epsilon(L)$ against band-limit $L$ for different parameters of the standard FB distribution in \figref{fig:spatial_error}, where it is evident that the spatial error converges to zero~(machine precision) as $L$ increases. Consequently, the standard FB distribution reconstructed from its coefficients converges to the formulation of standard FB distribution in spatial domain and thus validates the correctness of proposed analytic expression.

\subsection{Illustration - SFC Function}

\begin{figure}[tbp]
\centering
\includegraphics[width=0.8\columnwidth]{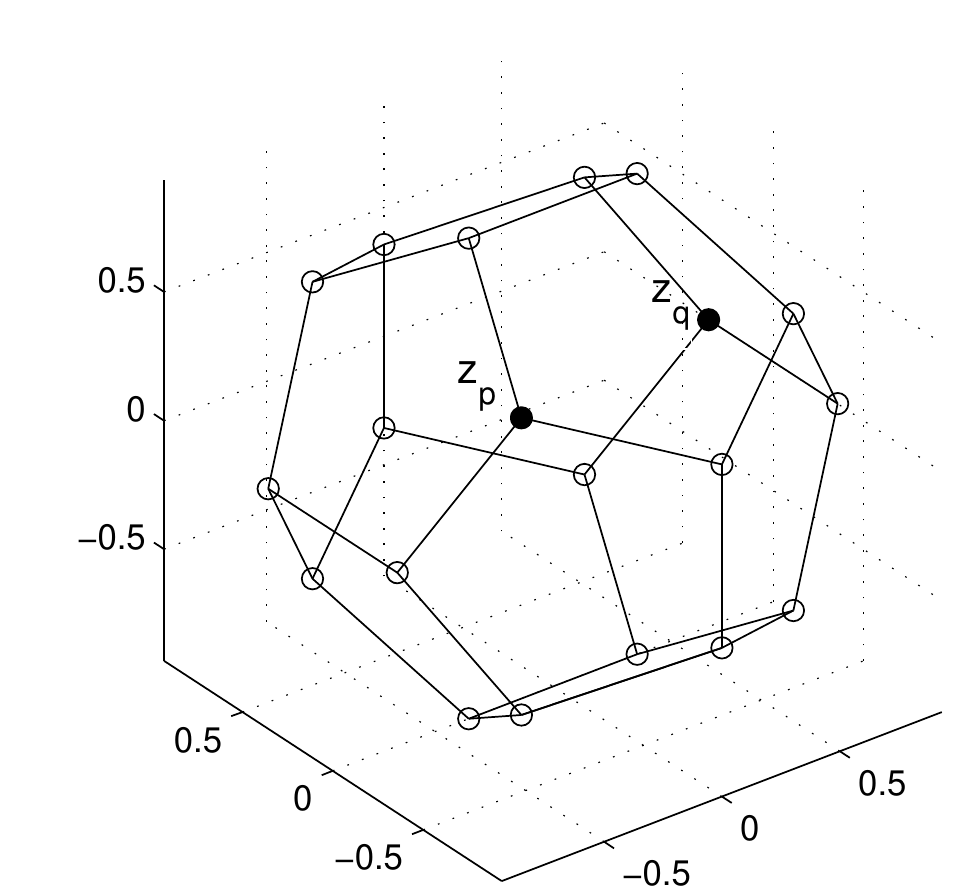}
\caption{Regular dodecahedron array (RDA) -- 20 antenna elements positioned at
the vertices of a regular dodecahedron inscribed in a sphere of radius $R=1$.}
\label{fig:rda_topology}
\end{figure}

Here, we validate the proposed closed-form expression for the SFC function
through numerical experiments. In our analysis, we consider both 2D and 3D
antenna array geometries in the form of uniform circular array (UCA) and
regular dodecahedron array (RDA), respectively. The antenna elements of
$M$-element UCA are placed at the following spatial positions
\begin{equation}
\bv{z}_p = \big[R\cos\frac{2\pi p}{M}, R\sin\frac{2\pi p}{M},0 \big]^T \in
\reals^{3},
\end{equation}
where $R$ denotes the circular radius of the array. For the RDA, 20 antenna
array elements are positioned at the vertices of a regular dodecahedron
inscribed in a sphere of radius $R$, as shown in \figref{fig:rda_topology} for
$R=1$. We assume that the AoA follows a standard Fisher-Bingham distribution.

\begin{figure}[tbp]
\centering
\includegraphics[width=0.75\columnwidth]{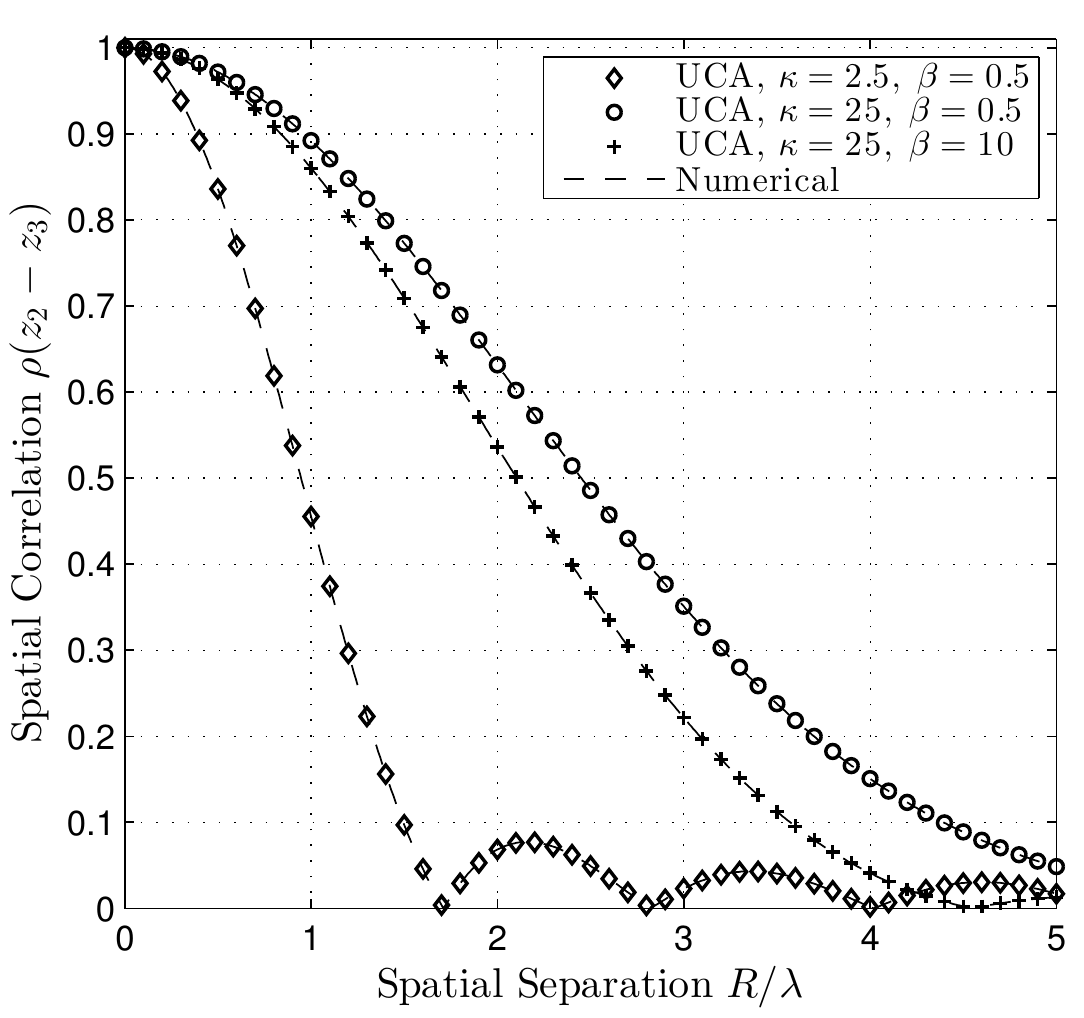}
\caption{Magnitude of the SFC function $\rho(\bv{z}_2 - \bv{z}_3)$ for
$16$-element UCA of radius $R$.}
\label{fig:spatcorr_UCA}
\end{figure}

Using the proposed closed-form expression in \eqref{Eq:sfc_analytic}, we
determine the SFC between the second and third UCA antenna elements and plot
the magnitude of the  SFC function $\rho(\bv{z } _2 - \bv{z}_3)$ in
\figref{fig:spatcorr_UCA} against the normalized radius $R/\lambda$. In the same
figure, we also plot the numerically evaluated SFC function, formulated in
\eqref{eq:spatcorr} and originally proposed in \cite{Mammasis:2009}, which
matches with the proposed closed-form expression for the SFC function. We
emphasise that numerical evaluation of the integrals employs computationally
intensive techniques to obtain sufficiently accurate results and is therefore
time consuming. Similarly, we compute the SFC between two antenna elements
positioned at $\bv{z}_p$ and $\bv{z}_q$ on RDA, which are indicated in
\figref{fig:rda_topology}, and plot the magnitude $\rho(\bv{z}_p-\bv{z}_q)$
\figref{fig:rda_topology}, which again matches with the numerically evaluated
SFC function given in \eqref{eq:spatcorr} and thus corroborates the correctness
of proposed SFC function.

\begin{figure}[tbp]
\centering
\includegraphics[width=0.75\columnwidth]{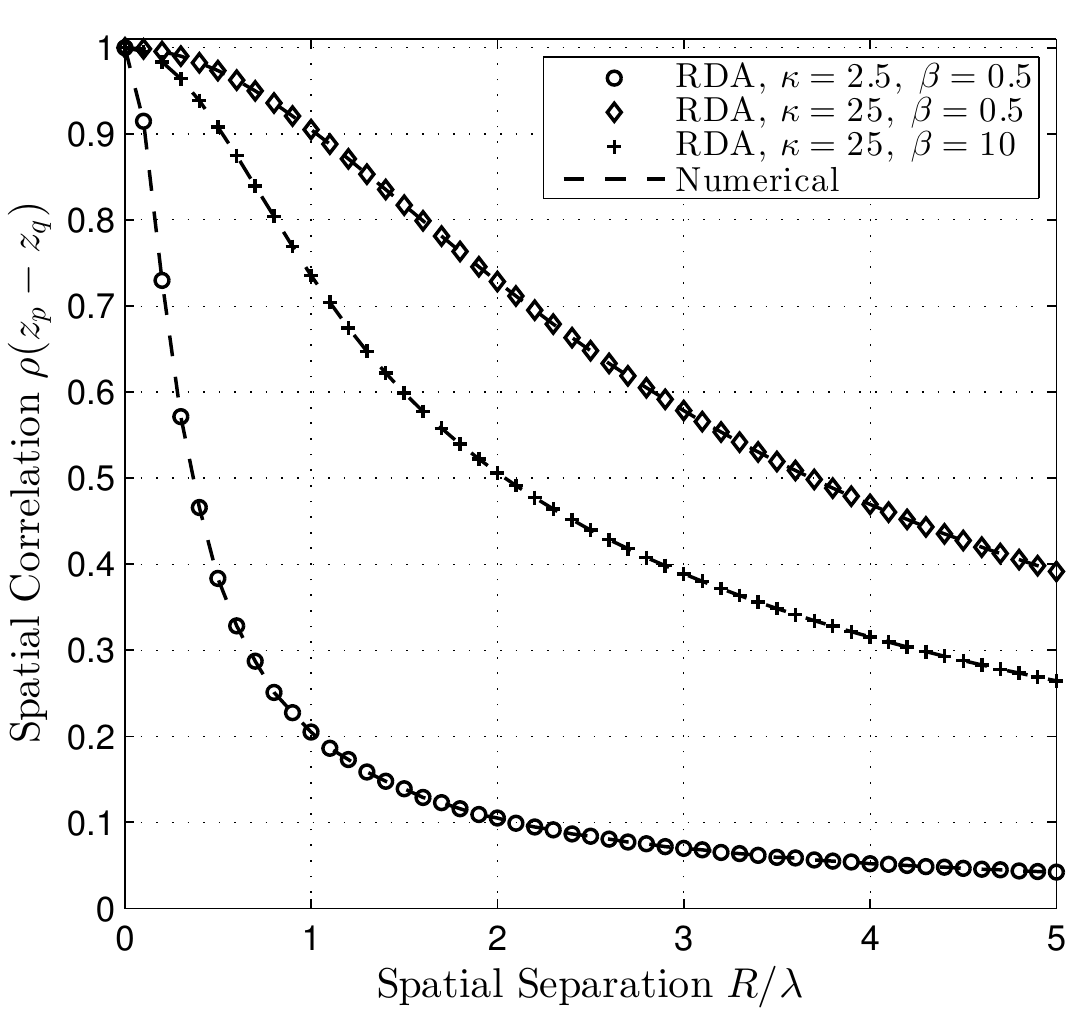}
\caption{Magnitude of the SFC function $\rho(\bv{z}_p - \bv{z}_q)$ between antenna elements, placed at vertices shaded in black in \figref{fig:rda_topology}, of
$20$-element RDA of radius $R$.}
\label{fig:spatcorr_RDA}
\end{figure}


\section{Conclusions} \label{sec:conclusion}
In this paper, the spherical harmonic expansion of the Fisher-Bingham
distribution and a closed-form expression that allows the
analytic computation of the spherical harmonic coefficients have been
presented. Using the expansion of plane waves in spherical harmonics and the
proposed closed-form expression for the spherical harmonic coefficients of the
FB5 distribution, we derived an analytic formula for the 3D SFC function between
two arbitrary points in 3D-space when the AoA of an incident signal has the
Fisher-Bingham distribution. Furthermore, we have validated the correctness of
proposed closed-form expressions for the spherical harmonic coefficients of the
Fisher-Bingham distribution and the 3D SFC using numerical experiments.
We have focussed on the use of Fisher-Bingham distribution for the computation
of SFC function. However, we believe that the proposed spherical harmonic
expansion of the Fisher-Bingham distribution has a great potential of
applicability in various applications for directional statistics and data
analysis on the sphere.

\vspace{10mm}
\bibliography{IEEEabrv,spatcorr}

\begin{thebibliography}{10}
\providecommand{\url}[1]{#1}
\csname url@samestyle\endcsname
\providecommand{\newblock}{\relax}
\providecommand{\bibinfo}[2]{#2}
\providecommand{\BIBentrySTDinterwordspacing}{\spaceskip=0pt\relax}
\providecommand{\BIBentryALTinterwordstretchfactor}{4}
\providecommand{\BIBentryALTinterwordspacing}{\spaceskip=\fontdimen2\font plus
\BIBentryALTinterwordstretchfactor\fontdimen3\font minus
  \fontdimen4\font\relax}
\providecommand{\BIBforeignlanguage}[2]{{%
\expandafter\ifx\csname l@#1\endcsname\relax
\typeout{** WARNING: IEEEtran.bst: No hyphenation pattern has been}%
\typeout{** loaded for the language `#1'. Using the pattern for}%
\typeout{** the default language instead.}%
\else
\language=\csname l@#1\endcsname
\fi
#2}}
\providecommand{\BIBdecl}{\relax}
\BIBdecl

\bibitem{Kent:1982}
J.~T. Kent, ``The {F}isher-{B}ingham distribution on the sphere,'' \emph{J. R.
  Statist. Soc.}, vol.~44, no.~1, pp. 71--80, Jun. 1982.

\bibitem{Leong:1998}
P.~Leong and S.~Carlile, ``Methods for spherical data analysis and
  visualization,'' \emph{J. Neuro. Metho.}, vol.~80, no.~2, pp. 191--200, 1998.

\bibitem{Langendijk:2001}
E.~H. Langendijk, D.~J. Kistler, and F.~L. Wightman, ``Sound localization in
  the presence of one or two distracters,'' \emph{J. Acoust. Soc. Am.}, vol.
  109, no.~5, pp. 2123--2134, 2001.

\bibitem{Peel:2001}
D.~Peel, W.~J. Whiten, and G.~J. McLachlan, ``Fitting mixtures of kent
  distributions to aid in joint set identification,'' \emph{J. American
  Statistical Asociation}, vol.~96, no. 453, pp. 56--63, 2001.

\bibitem{Kent:2005}
J.~T. Kent and T.~Hamelryck, ``Using the {F}isher-{B}ingham distribution in
  stochastic models for protein structure,'' \emph{In Barber S, Baxter P,
  Mardia K, Walls R, eds. Quantitative Biology, Shape ANalysis and Wavelets,
  Leeds University Press, Leeds, UK}, vol.~24, pp. 57--60, 2005.

\bibitem{Christou:2008}
C.~T. Christou, ``Beamforming spatially spread signals with the kent
  distribution,'' in \emph{Proc. IEEE Int. Conf. on Information Fusion}, 2008,
  pp. 1--7.

\bibitem{Lunga:2011}
D.~Lunga and O.~Ersoy, ``Kent mixture model for classification of remote
  sensing data on spherical manifolds,'' in \emph{Applied Imagery Pattern
  Recognition Workshop (AIPR), 2011 IEEE}.\hskip 1em plus 0.5em minus
  0.4em\relax IEEE, 2011, pp. 1--7.

\bibitem{Mammasis:2008}
K.~Mammasis and R.~W. Stewart, ``The {FB}5 distribution and its application in
  wireless communications,'' \emph{in Proc. of Int. ITG Workshop on Smart
  Antennas}, pp. 375--381, Feb. 2008.

\bibitem{Mammasis:2009}
K.~Mammasis and R.~W. Stewart, ``The {F}isher-{B}ingham spatial correlation
  model for multielement antenna systems,'' \emph{{IEEE} Trans. Veh. Technol.},
  vol.~58, no.~5, pp. 2130--2136, Jun. 2009.

\bibitem{Salz:1994}
J.~Salz and J.~H. Winters, ``Effect of fading correlation on adaptive arrays in
  digital mobile radio,'' \emph{{IEEE} Trans. Veh. Technol.}, vol.~43, no.~4,
  pp. 1049--1057, Nov. 1994.

\bibitem{Fang:2000}
L.~Fang, G.~Bi, and A.~C. Kot, ``New method of performance analysis for
  diversity reception with correlated rayleigh-fading signals,'' \emph{{IEEE}
  Trans. Veh. Technol.}, vol.~49, no.~5, pp. 1807--1812, 2000.

\bibitem{Shiu:2000}
D.~S. Shiu, G.~J. Foschini, M.~J. Gans, and J.~M. Kahn, ``Fading correlation
  and its effect on the capacity of multielement antenna systems,''
  \emph{{IEEE} Trans. Commun.}, vol.~48, no.~3, pp. 502--513, Mar. 2000.

\bibitem{Abdi:2002}
A.~Abdi and M.~Kaveh, ``A space-time correlation model for multielement antenna
  systems in mobile fading channels,'' \emph{{IEEE} J. Sel. Areas Commun.},
  vol.~20, no.~3, pp. 550--560, 2002.

\bibitem{Tsai:2004}
J.-A. Tsai, R.~M. Buehrer, and B.~D. Woerner, ``{BER} performance of a uniform
  circular array versus a uniform linear array in a mobile radio environment,''
  \emph{{IEEE} Trans. Wireless Commun.}, vol.~3, no.~3, pp. 695--700, 2004.

\bibitem{Kennedy:2007}
R.~A. Kennedy, P.~Sadeghi, T.~D. Abhayapala, and H.~M. Jones, ``Intrinsic
  limits of dimensionality and richness in random multipath fields,''
  \emph{{IEEE} Trans. Signal Process.}, vol.~55, no.~6, pp. 2542--2556, Jun.
  2007.

\bibitem{Kalkan:1997}
M.~Kalkan and R.~H. Clarke, ``Prediction of the space-frequency correlation
  function for base station diversity reception,'' \emph{{IEEE} Trans. Veh.
  Technol.}, vol.~46, no.~1, pp. 176--184, Feb. 1997.

\bibitem{Pedersen:1997}
K.~I. Pedersen, P.~E. Mogensen, and B.~H. Fleury, ``Power azimuth spectrum in
  outdoor environments,'' \emph{Electron. Lett.}, vol.~33, no.~18, pp.
  1583--1584, 1997.

\bibitem{Vaughan:1998}
R.~G. Vaughan, ``Pattern translation and rotation in uncorrelated source
  distributions for multiple beam antenna design,'' \emph{{IEEE} Trans.
  Antennas Propag.}, vol.~46, no.~7, pp. 982--990, Jul. 1998.

\bibitem{Abdi:2000}
A.~Abdi and M.~Kaveh, ``A versatile spatio-temporal correlation function for
  mobile fading channels with non-isotropic scattering,'' in \emph{Proc. Tenth
  IEEE Workshop on Statistical Signal and Array Processing}, Pocono Manor, PA,
  Aug. 2000, pp. 58--62.

\bibitem{Tsai:2002}
J.-A. Tsai, R.~Buehrer, and B.~Woerner, ``Spatial fading correlation function
  of circular antenna arrays with laplacian energy distribution,'' \emph{{IEEE}
  Commun. Lett.}, vol.~6, no.~5, pp. 178--180, May 2002.

\bibitem{Teal:2002}
P.~D. Teal, T.~D. Abhayapala, and R.~A. Kennedy, ``Spatial correlation for
  general distributions of scatterers,'' \emph{{IEEE} Signal Process. Lett.},
  vol.~9, no.~10, pp. 305--308, Oct. 2002.

\bibitem{Yong:2005}
S.~K. Yong and J.~Thompson, ``Three-dimensional spatial fading correlation
  models for compact mimo receivers,'' \emph{{IEEE} Trans. Wireless Commun.},
  vol.~4, no.~6, pp. 2856--2869, Nov. 2005.

\bibitem{Mammasis:2010}
K.~Mammasis and R.~W. Stewart, ``Spherical statistics and spatial correlation
  for multielement antenna systems,'' \emph{EURASIP Journal on Wireless
  Communication and Networking}, vol. 2010, Dec. 2010.

\bibitem{Lee:2012}
J.-H. Lee and C.-C. Cheng, ``Spatial correlation of multiple antenna arrays in
  wireless communication systems,'' \emph{Progress in Electromagnetics
  Research}, vol. 132, pp. 347--368, 2012.

\bibitem{Kennedy:2013b}
R.~A. Kennedy, Z.~Khalid, and Y.~F. Alem, ``Spatial correlation from multipath
  with {3D} power distributions having rotational symmetry,'' in \emph{Proc.
  Int. Conf. Signal Processing and Communication Systems, ICSPCS'2013}, Gold
  Coast, Australia, Dec. 2013, p.~7.

\bibitem{Kennedy-book:2013}
R.~A. Kennedy and P.~Sadeghi, \emph{Hilbert Space Methods in Signal
  Processing}.\hskip 1em plus 0.5em minus 0.4em\relax Cambridge, UK: Cambridge
  University Press, Mar. 2013.

\bibitem{Jeffrey:2007}
A.~Jeffrey and D.~Zwillinger, \emph{Table of integrals, series, and
  products}.\hskip 1em plus 0.5em minus 0.4em\relax Academic Press, 2007.

\bibitem{Trapani:2008}
S.~Trapani and J.~Navaza, ``{AMoRe}: Classical and modern,'' \emph{Acta
  Crystallogr. Sect. D}, vol.~64, no.~1, pp. 11--16, Jan. 2008.

\bibitem{Colton:2013}
D.~Colton and R.~Kress, \emph{Inverse Acoustic and Electromagnetic Scattering
  Theory}, 3rd~ed.\hskip 1em plus 0.5em minus 0.4em\relax New York, NY:
  Springer, 2013.

\bibitem{Jones:2002}
H.~M. Jones, R.~A. Kennedy, and T.~D. Abhayapala, ``On dimensionality of
  multipath fields: Spatial extent and richness,'' in \emph{Proc. IEEE Int.
  Conf. Acoustics, Speech, and Signal Processing, ICASSP'2002}, vol.~3,
  Orlando, FL, May 2002, pp. 2837--2840.

\bibitem{Khalid:2014}
Z.~Khalid, R.~A. Kennedy, and J.~D. McEwen, ``An optimal-dimensionality
  sampling scheme on the sphere with fast spherical harmonic transforms,''
  \emph{{IEEE} Trans. Signal Process.}, vol.~62, no.~17, pp. 4597--4610, Sep.
  2014.

\end{thebibliography}
\end{document}